\documentclass[a4paper,12pt]{article}

\usepackage{amsmath, amsthm, amsfonts, amssymb}

\usepackage{graphicx}

\theoremstyle{plain}
\newtheorem{theorem}{Theorem}[section]

\newtheorem{lemma}{Lemma}[section]

\theoremstyle{remark}
\newtheorem{remark}{Remark}[section]

\def\th{\theta}
\def\Om{\Omega}
\def\om{\omega}

\def\g{\gamma}
\def\G{\Gamma}
\def\l{\lambda}
\def\p{\partial}
\def\D{\Delta}

\def\E{\mbox{\rm e}}
\def\a{\alpha}
\def\b{\beta}
\def\si{\sigma}

\def\d{\delta}

\def\z{\zeta}
\def\vs{\varsigma}

\def\vp{\varphi}

\def\vr{\varrho}

\def\Odr{\mathcal{O}}
\def\H{W_2}

\def\Ho{W_{2,0}}

\def\di{\,\mathrm{d}}

\def\I{\mathrm{I}}
\def\iu{\mathrm{i}}

 \DeclareMathOperator{\RE}{Re}
 \DeclareMathOperator{\spec}{\sigma}
\DeclareMathOperator{\discspec}{\sigma_{disc}}
\DeclareMathOperator{\essspec}{\sigma_{ess}}

\DeclareMathOperator{\supp}{supp}

\DeclareMathOperator{\dvr}{div}
\DeclareMathOperator{\rank}{rank}
\DeclareMathOperator{\diag}{diag}

 \numberwithin{equation}{section}

\begin{document}
\allowdisplaybreaks

\title{Asymptotic behaviour of the spectrum of a
waveguide with distant perturbations}

\author{D.~Borisov}
\date{}
\maketitle

\begin{quote}
{\small { Nuclear Physics Institute, Academy of Sciences, 25068
\v Re\v z near Prague, Czechia
\\
Bashkir State Pedagogical University, October rev. st.~3a,
\\
 450000 Ufa, Russia
}
\\
E-mail: \texttt{borisovdi@yandex.ru}}
\end{quote}

\begin{quote}
{\small We consider the waveguide modelled by a $n$-dimensional
infinite tube. The operator we study is the Dirichlet Laplacian
perturbed by two distant perturbations. The perturbations are
described by arbitrary abstract operators ''localized'' in a
certain sense, and the distance between their ''supports'' tends
to infinity. We study the asymptotic behaviour of the discrete
spectrum of such system. The main results are a convergence
theorem and the asymptotics expansions for the eigenvalues. The
asymptotic behaviour of the associated eigenfunctions is
described as well. We also provide some particular examples of
the distant perturbations. The examples are the potential,
second order differential operator, magnetic Schr\"odinger
operator, curved and deformed waveguide, delta interaction, and
integral operator.}
\end{quote}

\section*{Introduction}

A multiple well problem for the Schr\"odinger operator attracted
much attention of many researches. A lot of works were devoted
to the studying of the semi-classical case (see, for instance,
\cite{KM}, \cite{CDS}, \cite{S}, and references therein). Quite
similar problems deal with the case when the small parameter at
higher derivative is replaced by a large distance between wells.
We mention papers \cite{H}, \cite{KM}, \cite{KS} as well as the
book \cite[Sec. 8.6]{D} devoted to such problems (see also
bibliography of these works). The main result of these works was
a description of the asymptotic behaviour of the eigenvalues and
the eigenfunctions as the distance between wells tended to
infinity. In \cite{HK} a double-well problem for the Dirac
operator with large distance between wells was studied. The
convergence and certain asymptotic results were established. We
also mention the paper \cite{KV}, where the usual potential was
replaced by a delta-potential supported by a curve. The results
of this paper imply the asymptotic estimate for the lowest
spectral gap in the case the curve consists of several disjoint
components and the distances between components tend to
infinity.

One of the ways of further developing of the mentioned problems
is to consider them not in $\mathbb{R}^n$, but in some other
unbounded domains. A good candidate is an infinite tube, since
such domain arises in the waveguide theory. One of such problems
has already been treated in \cite{BE}, where we considered a
quantum waveguide with two distant windows. The waveguide was a
two-dimensional infinite strip, where the Dirichlet Laplacian
was considered. The perturbation was two segments of the same
finite length on the boundary with Neumann boundary condition.
As the distance between the windows tended to infinity, a
convergence result and the asymptotic expansions for the
eigenvalues and the eigenfunctions were obtained. The technique
used in \cite{BE} employed essentially the symmetry of the
problem.

In the present paper we study a situation which is more general
in comparison with the articles cited. Namely, we deal with a
$n$-dimensional infinite tube where the Dirichlet Laplacian
$-\D^{(D)}$ is considered. The perturbation is two arbitrary
operators $\mathcal{L}_\pm$ ''localized'' in a certain sense.
The distance between their ''supports'' is assumed to be a large
parameter. In what follows we will call such perturbations as
distant perturbations. The precise description will be given in
the next section; we only say here that a lot of interesting
examples are particular cases of these operators (see Sec. 7).

Our main results are as follows. First we prove the convergence
result for the eigenvalues of the perturbed operator, and show
that the limiting values for these eigenvalues are the discrete
eigenvalues of the limiting operators
$-\D^{(D)}+\mathcal{L}_\pm$ and the threshold of the essential
spectrum. The most nontrivial result of the article is the
asymptotic expansions for the perturbed eigenvalues. Namely, we
obtain a scalar equation for these eigenvalues. Basing on this
equation, we construct the leading terms of the asymptotic
expansions for the perturbed eigenvalues. We also characterize
the asymptotic behaviour of the perturbed eigenfunctions.

If the distant perturbations are two same wells, it is
well-known that each limiting eigenvalue splits into pair of two
perturbed eigenvalues one of which being larger than a limiting
eigenvalue while the other being less. It is also known that the
leading terms of the asymptotics of these eigenvalues are
exponentially small w.r.t. the distance between well, and have
the same modulus but different signs. In the present article we
show that the similar phenomenon occur in the general situation
as well (see Theorem~\ref{th1.6}).

Let us also discuss the technique used in the paper. The kernel
of the approach is a scheme which allows us to reduce the
eigenvalue equation for the perturbed operator to an equivalent
operator equation in a special Hilbert space. The main advantage
of such reduction is that in the final equation the original
distant perturbations are replaced by an operator which is
meromorphic w.r.t. the spectral parameter and is multiplied by a
small parameter. We solve this equation explicitly by the
modification of the Birman-Schwinger technique suggested in
\cite{G1}, and in this way we obtain the described results. We
stress that our approach requires no symmetry restriction in
contrast to \cite{BE}. We should also say that the kind of
boundary condition is inessential, and for instance similar
problem for the Neumann Laplacian can be solved effectively,
too. Moreover, our approach can be applied to other problems
with distant perturbations not covered by the problem considered
here.

The article is organized as follows. In the next section we give
precise statement of the problem and formulate the main results.
In the second section we prove that the essential spectrum of
the perturbed operators is invariant w.r.t. the perturbations
and occupies a real semi-axis, while the discrete spectrum
contains finitely many eigenvalues. The third section is devoted
to the studying of the limiting operators; we collect there some
preliminary facts required for the proof of the main results. In
the fourth section we provide the aforementioned scheme
transforming the original perturbed eigenvalue equation to an
equivalent operator equation. Then we solve this equation
explicitly. The fifth section is devoted to the proof of the
convergence result. The asymptotic formulas for the perturbed
eigenelements are established in the sixth section. The final
seventh section contains some examples of the operators
$\mathcal{L}_\pm$.

\section{Statement of the problem and formulation of the
results}

Let $x=(x_1,x')$ and $x'=(x_2,\ldots,x_n)$ be Cartesian
coordinates in $\mathbb{R}^n$ and $\mathbb{R}^{n-1}$,
respectively, and let $\om$ be a bounded domain in
$\mathbb{R}^{n-1}$ having infinitely differentiable boundary. We
assume that $n\geqslant2$. By $\Pi$ we denote an infinite tube
$\mathbb{R}\times\om$. Given any bounded domain $Q\subset\Pi$,
by $L_2(\Pi,Q)$ we denote the subset of the functions from
$L_2(\Pi)$ whose support lies inside $\overline{Q}$. For any
domain $\Om\subseteq\Pi$ and $(n-1)$-dimensional manifold
$S\subset\overline{\Om}$ the symbol $\Ho^j(\Om,S)$ will indicate
the subset of the functions from $\H^j(\Om)$ vanishing on $S$.
If $S=\p\Om$, we will write shortly $\Ho^j(\Om)$.

Let $\Om_\pm$ be a pair of bounded subdomains of $\Pi$ defined
as $\Om_\pm:=(-a_\pm,a_\pm)\times\om$, where
$a_\pm\in\mathbb{R}$ are fixed positive numbers. We let
$\g_\pm:=(-a_\pm,a_\pm)\times\p\om$. By $\mathcal{L}_\pm$ we
denote a pair of bounded linear operators from
$\Ho^j(\Om_\pm,\g_\pm)$ into $L_2(\Pi,\Om_\pm)$. We assume that
for all $u_1,u_2\in\Ho^j(\Om_\pm,\g_\pm)$ the identity
\begin{equation}\label{1.1}
(\mathcal{L}_\pm u_1,u_2)_{L_2(\Om_\pm)}=(u_1,\mathcal{L}_\pm
u_2)_{L_2(\Om_\pm)}
\end{equation}
holds true. We also suppose that the operators $\mathcal{L}_\pm$
satisfy the estimate
\begin{equation}\label{1.2}
(\mathcal{L}_\pm u,u)_{L_2(\Om_\pm)}\geqslant -c_0\|\nabla
u\|_{L_2(\Om_\pm)}^2-c_1\|u\|_{L_2(\Om_\pm)}^2
\end{equation}
for all $u\in \Ho^2(\Om_\pm,\g_\pm)$, where the constants $c_0$,
$c_1$ are independent of $u$, and
\begin{equation}\label{1.3}
c_0<1.
\end{equation}
Since the restriction of each function $u\in\Ho^2(\Pi)$ on
$\Om_\pm$ belongs to $\Ho^2(\Om_\pm,\g_\pm)$, we can also regard
the operators $\mathcal{L}_\pm$ as unbounded ones in $L_2(\Pi)$
with the domain $\Ho^2(\Pi)$.

By $\mathcal{S}(a)$ we denote a shift operator in $L_2(\Pi)$
acting as $(\mathcal{S}(a)u)(x):=u(x_1+a,x')$, and for any $l>0$
we introduce the operator
\begin{equation*}
\mathcal{L}_l:=\mathcal{S}(l)\mathcal{L}_-\mathcal{S}(-l)+
\mathcal{S}(-l)\mathcal{L}_+\mathcal{S}(l).
\end{equation*}
Clearly, the operator $\mathcal{L}_l$ depends on values its
argument takes on the set $\{x: (x_1+l,x')\in\Om_-\}\cup\{x:
(x_1-l,x')\in\Om_+\}$. As $l\to+\infty$, this set consists of
two components separated by the distance $2l$. This is why we
can regard the operator $\mathcal{L}_l$ as the distant
perturbations formed by $\mathcal{L}_-$ and $\mathcal{L}_+$.

The main object of our study is the spectrum of the operator
$\mathcal{H}_l:=-\D^{(D)}+\mathcal{L}_l$ in $L_2(\Pi)$ with
domain $\Ho^2(\Pi)$. Here $-\D^{(D)}$ indicates the Laplacian in
$L_2(\Pi)$ with the domain $\Ho^2(\Pi)$.

We suppose that the operators $\mathcal{L}_\pm$ are so that the
operator $\mathcal{H}_l$ is self-adjoint. The main aim of this
paper is to study the behaviour of the spectrum of
$\mathcal{H}_l$ as $l\to+\infty$.

In order to formulate the main results we need to introduce
additional notations. We will employ the symbols $\spec(\cdot)$,
$\essspec(\cdot)$, $\discspec(\cdot)$ to indicate the spectrum,
the essential and discrete one of an operator. We denote
$\mathcal{H}_\pm:=-\D^{(D)}+\mathcal{L}_\pm$, and suppose that
these operators with domain $\Ho^2(\Pi)$ are self-adjoint in
$L_2(\Pi)$.

\begin{remark}\label{rm1.0}
We note that the assumptions (\ref{1.1}), (\ref{1.2}),
(\ref{1.3}) do not imply the self-adjointness of $\mathcal{H}_l$
and $\mathcal{H}_\pm$, and we can employ here neither KLMN
theorem no Kato-Rellich theorem. At the same time, if the
operators $\mathcal{L}_\pm$ satisfy stricter assumption and are
$-\D^{(D)}$-bounded with the bound less than one, it implies the
self-adjointness of $\mathcal{H}_l$ and $\mathcal{H}_\pm$.
\end{remark}

Let $\nu_1>0$ be the ground state of the Dirichlet Laplacian in
$\om$.

Our first result is
\begin{theorem}\label{th1.1}
The essential spectra of the operators $\mathcal{H}_l$,
$\mathcal{H}_+$, $\mathcal{H}_-$ coincide with the semi-axis
$[\nu_1,+\infty)$. The discrete spectra of the operator
$\mathcal{H}_l$, $\mathcal{H}_+$, $\mathcal{H}_-$ consist of
finitely many real eigenvalues.
\end{theorem}

We denote
$\si_*:=\discspec(\mathcal{H}_-)\cup\discspec(\mathcal{H}_+)$.
Let $\l_*\in\si_*$ be a $p_-$-multiple eigenvalue of
$\mathcal{H}_-$ and $p_+$-multiple eigenvalue of
$\mathcal{H}_+$, where $p_\pm$ is taken being zero if
$\l_*\not\in\discspec(\mathcal{H}_\pm)$. In this case we will
say that $\l_*$ is $(p_-+p_+)$-multiple.

\begin{theorem}\label{th1.2}
Each discrete eigenvalue of $\mathcal{H}_l$ converges to one of
the numbers from $\si_*$ or to $\nu_1$ as $l\to+\infty$.
\end{theorem}

\begin{theorem}\label{th1.3}
If $\l_*\in\si_*$ is $(p_-+p_+)$-multiple, the total
multiplicity of the eigenvalues of $\mathcal{H}_l$ converging to
$\l_*$ equals $p_-+p_+$.
\end{theorem}

In what follows we will employ symbols $(\cdot,\cdot)_X$ and
$\|\cdot\|_X$ to indicate the inner product and the norm in a
Hilbert space $X$.

Suppose that $\l_*\in\si_*$ is $(p_-+p_+)$-multiple, and
$\psi_i^\pm$, $i=1,\ldots,p_\pm$, are the eigenfunctions of
$\mathcal{H}_\pm$ associated with $\l_*$ and orthonormalized in
$L_2(\Pi)$. If $p_-\geqslant 1$, we denote
\begin{align*}
\boldsymbol{\phi}_i(\cdot,l)&:=(0,\mathcal{L}_+\mathcal{S}(2l)
\psi_i^-)\in L_2(\Om_-)\oplus L_2(\Om_+),
\\
\mathcal{T}_1^{(i)}\boldsymbol{f}&:=(f_-,\psi_i^-)_{L_2(\Om_-)},\quad
i=1,\ldots,p_-,
\end{align*}
where $\boldsymbol{f}:=(f_-,f_+)\in L_2(\Om_-)\oplus
L_2(\Om_+)$. If $p_+\geqslant 1$, we denote
\begin{align*}
\boldsymbol{\phi}_{i+p_-}(\cdot,l)&:=(\mathcal{L}_-\mathcal{S}(-2l)
\psi_i^+,0)\in L_2(\Om_-)\oplus L_2(\Om_+),
\\
\mathcal{T}_1^{(i+p_-)}\boldsymbol{f}&:=
(f_+,\psi_i^+)_{L_2(\Om_+)},\quad i=1,\ldots,p_+.
\end{align*}
In the fourth section we will show that the operator
\begin{equation}\label{1.25}
\mathcal{T}_2(\l,l)\boldsymbol{f}:=\big(\mathcal{L}_-
\mathcal{S}(-2l)(\mathcal{H}_+-\l)^{-1}f_+,\mathcal{L}_+
\mathcal{S}(2l) (\mathcal{H}_--\l)^{-1}f_-\big)
\end{equation}
in $L_2(\Om_-)\oplus L_2(\Om_+)$ satisfies the relation
\begin{equation}\label{4.11}
\mathcal{T}_2(\l,l)=-\frac{1}{\l-\l_*}\sum\limits_{i=1}^{p}
\boldsymbol{\phi}_i(\cdot,l)\mathcal{T}_1^{(i)}+
\mathcal{T}_3(\l,l),
\end{equation}
for $\l$ close to $\l_*$, where $p:=p_-+p_+$, and the norm of
$\mathcal{T}_3$ tends to zero as $l\to+\infty$ uniformly in $l$.
We introduce the matrix
\begin{equation*}
\mathrm{A}(\l,l):=
\begin{pmatrix}
A_{11}(\l,l)& \ldots& A_{1p}(\l,l)
\\
\vdots& &\vdots
\\
A_{p1}(\l,l)&\ldots&A_{pp}(\l,l)
\end{pmatrix},
\end{equation*}
where
$A_{ij}(\l,l):=\mathcal{T}_1^{(i)}(\I+\mathcal{T}_3(\l,l))^{-1}
\boldsymbol{\phi}_j(\cdot,l)$.

\begin{theorem}\label{th1.4}
Let $\l_*\in\si_*$ be $(p_-+p_+)$-multiple, and let
$\l_i=\l_i(l)\xrightarrow[l\to+\infty]{}\l_*$, $i=1,\ldots,p$,
$p:=p_-+p_+$, be the eigenvalues of $\mathcal{H}_l$ taken
counting multiplicity and ordered as follows
\begin{equation}\label{1.4}
0\leqslant |\l_1(l)-\l_*|\leqslant |\l_2(l)-\l_*|\leqslant\ldots
\leqslant |\l_p(l)-\l_*|.
\end{equation}
These eigenvalues solve the equation
\begin{equation}\label{4.15}
\det\big((\l-\l_*)\mathrm{E}-\mathrm{A}(\l,l)\big)=0,
\end{equation}
and satisfy the asymptotic formulas
\begin{equation}\label{1.5}
\l_i(l)=\l_*+\tau_i(l)\left(1+\Odr\left(l^{\frac{2}{p}}
\E^{-\frac{4l}{p}\sqrt{\nu_1-\l_*}}\right)\right),\quad
l\to+\infty.
\end{equation}
Here
\begin{equation}\label{1.6}
\tau_i=\tau_i(l)=\Odr(\E^{-2l\sqrt{\nu_1-\l_*}}),\quad
l\to+\infty,
\end{equation}
are the zeroes of the polynomial $\det\big(\tau
\mathrm{E}-\mathrm{A}(\l_*,l)\big)$ taken counting multiplicity
and ordered as follows
\begin{equation}\label{1.7}
0\leqslant |\tau_1(l)|\leqslant |\tau_2(l)|\leqslant\ldots
\leqslant |\tau_p(l)|.
\end{equation}
The eigenfunctions associated with $\l_i$ satisfy the asymptotic
representation
\begin{equation}\label{1.8}
\psi_i(x,l)=\sum\limits_{i=1}^{p_-} k_{i,j}\psi_j^-(x_1+l,x')+
\sum\limits_{i=1}^{p_+} k_{i,j+p_-}\psi_j^+(x_1-l,x')+
\Odr(\E^{-2l\sqrt{\nu_1-\l_*}}),\quad l\to+\infty,
\end{equation}
in $\H^2(\Pi)$-norm. The numbers $k_{i,j}$ are the components of
the vectors
\begin{equation*}
\boldsymbol{k}_i=\boldsymbol{k}_i(l)= \left(k_{i,1}(l) \ldots
k_{i,p}(l)\right)^{t}
\end{equation*}
solving the system
\begin{equation}
\big((\l-\l_*)\mathrm{E}-\mathrm{A}(\l,l)\big)\boldsymbol{k}=0,
\label{4.14}
\end{equation}
for $\l=\l_i(l)$, and satisfying the condition
\begin{equation}\label{5.8}
(\boldsymbol{k}_i,\boldsymbol{k}_j)_{\mathbb{C}^p}=
\begin{cases}
1,& \text{if}\quad i=j,
\\
\Odr(l\E^{-2l\sqrt{\nu_1-\l_*}}),&\text{if}\quad i\not=j.
\end{cases}
\end{equation}
\end{theorem}

According to this theorem, the leading terms of the asymptotics
expansions for the eigenvalues $\l_i$ are determined by the
matrix $\mathrm{A}(\l_*,l)$. At the same time, in applications
it could be quiet complicated to calculate this matrix
explicitly. This is why in the following theorems we provide one
more way of calculating the asymptotics expansions.

We will say that a square matrix $\mathrm{A}(l)$ satisfies the
condition (A), if it is diagonalizable and the determinant of
the matrix formed by the normalized eigenvectors of
$\mathrm{A}(l)$ is separated from zero uniformly in $l$ large
enough.

\begin{theorem}\label{th1.5}
Let the hypothesis of Theorem~\ref{th1.4} hold true. Suppose
that the matrix $\mathrm{A}(\l_*,l)$  can be represented as
\begin{equation}\label{1.10}
\mathrm{A}(\l_*,l)=\mathrm{A}_0(l)+\mathrm{A}_1(l),
\end{equation}
where the matrix $\mathrm{A}_0$ satisfies the condition (A), and
$\|\mathrm{A}_1(l)\|\to0$ as $l\to+\infty$. In this case the
eigenvalues $\l_i$ of $\mathcal{H}_l$ satisfy the asymptotic
formulas
\begin{equation}\label{1.11}
\l_i=\l_*+\tau_i^{(0)}\big(1+\Odr(l^2\E^{-4l
\sqrt{\nu_1-\l_*}})\big)+ \Odr(\|\mathrm{A}_1(l)\|),\quad
l\to+\infty.
\end{equation}
Here $\tau_i^{(0)}=\tau_i^{(0)}(l)$ are the roots of the
polynomial $\det\big(\tau\mathrm{E}-\mathrm{A}_0(l)\big)$ taken
counting multiplicity and ordered as follows
\begin{equation*}
0\leqslant |\tau_1^{(0)}(l)|\leqslant
|\tau_2^{(0)}(l)|\leqslant\ldots \leqslant |\tau_p^{(0)}(l)|.
\end{equation*}
Each of these roots satisfies the estimate
\begin{equation}\label{1.13}
\tau_i^{(0)}(l)=\Odr(\|\mathrm{A}_0(l)\|),\quad l\to+\infty.
\end{equation}
\end{theorem}

This theorem states that the leading terms of the asymptotics
for the eigenvalues can be determined by that of the asymptotics
for $A(\l_*,l)$. At the same time, the estimate for the error
term in (\ref{1.11}) can be worse than that in (\ref{1.5}). In
the following theorem we apply Theorem~\ref{th1.5} to several
important particular cases.

Let $\nu_2>\nu_1$ be the second eigenvalue of the negative
Dirichlet Laplacian in $\om$, and $\phi_1=\phi_1(x')$ be the
eigenvalue associated with $\nu_1$ and normalized in $L_2(\om)$.
In the fifth section we will prove
\begin{lemma}\label{lm1.1}
Let the hypothesis of Theorem~\ref{th1.4} hold true, and
$p_\pm\geqslant 1$. Then the functions $\psi_i^\pm$ can be
chosen so that
\begin{equation}\label{1.14}
\psi_1^\pm(x)=\b_\pm\E^{\pm\sqrt{\nu_1-\l_*}x_1}\phi_1(x')+
\Odr(\E^{\pm\sqrt{\nu_2-\l_*}x_1}),\quad
\psi_i^\pm(x)=\Odr(\E^{\pm\sqrt{\nu_2-\l_*}x_1}),
\end{equation}
as $x_1\to\mp\infty$,  $i=2,\ldots,p_\pm$, and the functions
$\psi_i^\pm$ are orthonormalized in $L_2(\Pi)$.
\end{lemma}

\begin{theorem}\label{th1.7}
Let the hypothesis of Theorem~\ref{th1.5} hold true, and
$p_+=0$. Then the eigenvalues $\l_i$ satisfy the asymptotic
formulas
\begin{equation}\label{1.18}
\begin{aligned}
&\l_i(l)=\l_*+\Odr\big(\E^{-2l(\sqrt{\nu_1-\l_*}+
\sqrt{\nu_2-\l_*})}\big),\quad i=1,\ldots,p-1,
\\
&\l_p(l)=\l_*-2\sqrt{\nu_1-\l_*}|\b_-|^2\widetilde{\b}_-
\E^{-4l\sqrt{\nu_1-\l_*}}+\Odr\big(\E^{-2l(\sqrt{\nu_1-\l_*}
+\sqrt{\nu_2-\l_*})}+l^2\E^{-6l\sqrt{\nu_1-\l_*}}\big),
\end{aligned}
\end{equation}
where the constant $\widetilde{\b}_-$ is  determined uniquely by
the identity
\begin{equation}\label{1.19}
\begin{aligned}
&U_+(x)=\widetilde{\b}_-\E^{-\sqrt{\nu_1-\l_*}x_1}\phi_1(x')+
\Odr(\E^{\sqrt{\nu_2-\l_*}x_1}),\quad x_1\to-\infty,
\\
&U_+:=(\mathcal{H}_+-\l_*)^{-1} \mathcal{
L}_+\big(\E^{-\sqrt{\nu_1-\l_*}x_1}\phi_1(x')\big).
\end{aligned}
\end{equation}
\end{theorem}

\begin{theorem}\label{th1.8}
Let the hypothesis of Theorem~\ref{th1.5} hold true, and
$p_-=0$. Then the eigenvalues $\l_i$ satisfy the asymptotic
formulas
\begin{equation}\label{1.20}
\begin{aligned}
&\l_i(l)=\l_*+\Odr\big(\E^{-2l(\sqrt{\nu_1-\l_*}+
\sqrt{\nu_2-\l_*})}\big),\quad i=1,\ldots,p-1,
\\
&\l_p(l)=\l_*-2\sqrt{\nu_1-\l_*}|\b_+|^2\widetilde{\b}_+
\E^{-4l\sqrt{\nu_1-\l_*}}+\Odr\big(\E^{-2l(\sqrt{\nu_1-\l_*}
+\sqrt{\nu_2-\l_*})}+l^2\E^{-6l\sqrt{\nu_1-\l_*}}\big),
\end{aligned}
\end{equation}
where the constant $\widetilde{\b}_+$ is uniquely determined by
the identity
\begin{align*}
&U_+(x)=\widetilde{\b}_+\E^{\sqrt{\nu_1-\l_*}x_1}\phi_1(x')+
\Odr(\E^{\sqrt{\nu_2-\l_*}x_1}),\quad x_1\to+\infty,
\\
&U_+:=(\mathcal{H}_--\l_*)^{-1} \mathcal{
L}_+\big(\E^{-\sqrt{\nu_1-\l_*}x_1}\phi_1(x')\big)
\end{align*}
\end{theorem}

These two theorems treat the first possible case when the number
$\l_*\in\sigma$ is the eigenvalue of one of the operators
$\mathcal{H}_\pm$ only. The formulas (\ref{1.18}), (\ref{1.20})
give the asymptotic expansion for the eigenvalue $\l_p$, and the
asymptotic estimates for the other eigenvalues. At the same
time, given arbitrary $\mathcal{L}_\pm$ and an eigenvalue $\l_*$
of $\mathcal{H}_\pm$, this eigenvalue is simple. In this case
$p=1$, and by Theorem~\ref{th1.3} there exists the unique
perturbed eigenvalue converging to $\l_*$, and
Theorems~\ref{th1.7},~\ref{th1.8} provide its asymptotics.

\begin{theorem}\label{th1.6}
Let the hypothesis of Theorem~\ref{th1.4} hold true, and
$p_\pm\geqslant 1$. Then the eigenvalues $\l_i$ satisfy the
asymptotic formulas
\begin{equation}\label{1.17}
\begin{aligned}
&\l_i(l)=\l_*+\Odr(l\E^{-4l\sqrt{\nu_1-\l_*}}),\quad
i=1,\ldots,p-2,
\\
&\l_{p-1}(l)=\l_*-2|\b_-\b_+|\sqrt{\nu_1-\l_*}\E^{-2l
\sqrt{\nu_1-\l_*}}+\Odr(l\E^{-4l\sqrt{\nu_1-\l_*}}),
\\
&\l_{p}(l)=\l_*+2|\b_-\b_+|\sqrt{\nu_1-\l_*}\E^{-2l
\sqrt{\nu_1-\l_*}}+\Odr(l\E^{-4l\sqrt{\nu_1-\l_*}}),
\end{aligned}
\end{equation}
as $l\to+\infty$.
\end{theorem}

This theorem deals with the second possible case when the number
$\l_*\in\si$ is an eigenvalue of both operators
$\mathcal{H}_\pm$. Similarly to
Theorems~\ref{th1.7},~\ref{th1.8}, the formulas (\ref{1.17})
give the asymptotic expansions for $\l_{p-1}$ and $\l_p$, and
the asymptotic estimates for the other eigenvalues. At the same
time, the most possible case is that $\l_*$ is a simple
eigenvalue of $\mathcal{H}_+$ and $\mathcal{H}_-$. In this case
there exist only two perturbed eigenvalues converging to $\l_*$,
and Theorem~\ref{th1.6} give their asymptotic expansions.

Suppose now that under the hypothesis of Theorem~\ref{th1.6} the
inequality $\b_-\b_+\not=0$  holds true. In this case the
leading terms of the asymptotic expansions for $\l_{p-1}$ and
$\l_p$ have the same modulus but different signs. Moreover,
these eigenvalues are simple. This situation is similar to what
one has when dealing with a double-well problem with symmetric
wells. We should also stress that in our case we assume no
symmetry condition for the distant perturbations. It means that
the mentioned phenomenon is the general situation and not the
consequence of the symmetry. We also observe that the formulas
(\ref{1.17}) allow us to estimate the spectral gap between
$\l_{p-1}$ and $\l_p$,
\begin{equation*}
\l_2(l)-\l_1(l)=4|\b_-\b_+|\sqrt{\nu_1-\l_*}\E^{-2l
\sqrt{\nu_1-\l_*}}+\Odr(l\E^{-4l\sqrt{\nu_1-\l_*}}),\quad
l\to+\infty.
\end{equation*}

In conclusion we note that it is possible to calculate the
asymptotic expansions for the eigenvalues $\l_i$, $i\leqslant
p-1$, in Theorem~\ref{th1.7},~\ref{th1.8}, and for $\l_i$,
$i\leqslant p-2$, in Theorem~\ref{th1.6}. In order to do it, one
should employ the technique of the proofs of the mentioned
theorems and extract the next-to-leading term of the asymptotic
for $A(\l_*,l)$. Then the obtained terms of this asymptotics
should be treated as the matrix $A_0(l)$ in (\ref{1.10}). We do
not adduce such calculations in the article in order not to
overload the text by quite bulky and technical details.

\section{Proof of Theorem~\ref{th1.1}}

Let $\Om$ be a bounded non-empty subdomain of $\Pi$ defined as
$\Om:=(-a,a)\times\om$, where $a\in\mathbb{R}$, $a>0$,
$\g:=(-a,a)\times\p\om$, and let $\mathcal{L}$ be an arbitrary
bounded operator from $\Ho^2(\Om,\g)$ into $L_2(\Pi,\Om)$.
Assume that for all $u,u_1,u_2\in\Ho^2(\Om,\g)$ the relations
\begin{equation}\label{2.1}
\begin{aligned}
&(\mathcal{L} u_1,u_2)_{L_2(\Om)}=(u_1,\mathcal{L}
u_2)_{L_2(\Om)}, \quad (\mathcal{L} u,u)_{L_2(\Om)}\geqslant
-c_0\|\nabla u\|_{L_2(\Om)}^2-c_1\|u\|_{L_2(\Om)}^2
\end{aligned}
\end{equation}
hold true, where $c_0$, $c_1$ are constants, and $c_0$ obeys
(\ref{1.3}). As in the case of the operators $\mathcal{L}_\pm$,
we can also regard the operator $\mathcal{L}$ as an unbounded
one in $L_2(\Pi)$ with domain $\Ho^2(\Pi)$. Suppose also that
the operator $\mathcal{H}_{\mathcal{L}}:=-\D^{(D)}+\mathcal{L}$
with domain $\Ho^2(\Pi)$ is self-adjoint in $L_2(\Pi)$.

\begin{lemma}\label{lm2.1}
$\essspec(\mathcal{H}_{\mathcal{L}})=[\nu_1,+\infty)$.
\end{lemma}
\begin{proof}
We employ Weyl criterion to prove the theorem. Let
$\l\in[\nu_1,+\infty)$ and let $\chi=\chi(t)\in
C^\infty(\mathbb{R})$ be a  cut-off function equalling one as
$t<0$, vanishing as $t>1$, and
\begin{equation*}
\int\limits_{0}^{1}\chi^2(t)\di t=\frac{1}{2}.
\end{equation*}
We introduce a sequence of the functions
\begin{equation*}
u_p(x):=\frac{\E^{\iu\sqrt{\l-\nu_1}x_1}}{\sqrt{(2p+1)|\om|}}
\chi(|x_1|-p)\phi_1(x').
\end{equation*}
It is easy to make sure that
\begin{equation}\label{2.3a}
\|u_p\|_{L_2(\Pi)}=1,\quad
(u_p,\z)_{L_2(\Pi)}\xrightarrow[p\to+\infty]{}0
\end{equation}
for each function $\z\in C^\infty(\Pi)$ vanishing on $\p\Pi$ and
having a compact support. Hence, $u_p$ converges to zero weakly
in $L_2(\Pi)$ as $p\to+\infty$. Moreover,
$\|u_p\|_{\H^2(\Om)}\to0$  as $p\to+\infty$. It implies
immediately that $\|\mathcal{L}u_p\|_{L_2(\Om)}\to0$ as
$p\to+\infty$. Thus,
\begin{equation*}
\|(-\D^{(D)}+\mathcal{L}-\l)u_p\|_{L_2(\Pi)}\to0,\quad
p\to+\infty.
\end{equation*}
By (\ref{2.3a}) it yields that $u_p$ is a singular sequence for
$\mathcal{H}_{\mathcal{L}}$ at $\l$. Therefore,
$[\nu_1,+\infty)\subseteq\essspec(\mathcal{H}_{\mathcal{L}})$.
Let us prove the opposite inclusion.

Suppose $\l\in\essspec(\mathcal{H}_{\mathcal{L}})$. By Weyl
criterion it follows that there exists a singular sequence $u_p$
for  $\mathcal{H}_{\mathcal{L}}$ at $\l$. Employing $u_p$,  we
are going to construct a singular sequence for $-\D^{(D)}$ at
$\l$. Due to the obvious identity
$\essspec(-\D^{(D)})=[\nu_1,+\infty)$, such fact will complete
the proof.

Without loss of generality we assume that
$\|u_p\|_{L_2(\Pi)}=1$. Employing the inequality in (\ref{2.1})
and denoting $f_p:=\mathcal{H}_{\mathcal{L}}u_p$, we obtain a
chain of relations,
\begin{align*}
&(f_p,u_p)_{L_2(\Pi)}=(\mathcal{H}_{\mathcal{L}}u_p,u_p)_{L_2(\Pi)}
=\|\nabla u_p\|_{L_2(\Pi)}^2-\l+(\mathcal{L}u_p,u_p)_{L_2(\Pi)},
\\
&\|\nabla u_p\|_{L_2(\Pi)}^2=\l-(\mathcal{L}u_p,u_p)_{L_2(\Pi)}+
(f_p,u_p)_{L_2(\Pi)}\leqslant \l+c_0\|\nabla
u_p\|_{L_2(\Pi)}^2+c_1+\|f_p\|_{L_2(\Pi)},
\\
&(1-c_0)\|\nabla u_p\|_{L_2(\Pi)}^2\leqslant
\l+c_1+\|f_p\|_{L_2(\Pi)}\leqslant C,
\end{align*}
where the constant $C$ is independent of $p$. Therefore, $u_p$
is uniformly (in $p$) bounded in the norm of $\H^1(\Pi)$.

Since the operator $\mathcal{H}_{\mathcal{L}}$ is self-adjoint,
the inverse operator $(\mathcal{H}_{\mathcal{L}}-\iu)^{-1}$
exists and is bounded as an operator in $L_2(\Pi)$. The range of
the inverse operator is $\Ho^2(\Pi)$ and we can thus regard it
as an operator from $L_2(\Pi)$ into $\Ho^2(\Pi)$. The operator
$(\mathcal{H}_\mathcal{L}-\iu) : \Ho^2(\Pi)\to L_2(\Pi)$ is
bounded.  Therefore by Banach theorem on inverse operator (see,
for instance, \cite[Ch. 6, Sec. 23.1, Theorem 2]{KF}) the
operator $(\mathcal{H}_{\mathcal{L}}-\iu)^{-1}: L_2(\Pi)\to
\Ho^2(\Pi)$ is bounded. This fact, the obvious identity
$(\mathcal{H}_{\mathcal{L}}-\iu)u_p=(\l-\iu)u_p+f_p$ and the
properties of $u_p$ and $f_p$ imply that the functions $u_p$ are
bounded in $\H^2(\Pi)$-norm uniformly in $p$. Extracting if
needed a subsequence from $\{u_p\}$, we can assume that this
sequence converges to zero weakly in $\Ho^2(\widetilde{\Om})$
and strongly in $\H^1(\widetilde{\Om})$, where
$\widetilde{\Om}:=(-a-1,a+1)\times\om$.

We denote  $v_p(x):=\big(1-\chi(|x_1|-a)\big)u_p(x)$. Clearly,
$v_p=0$ as $x\in\Om$. Hence,
\begin{equation*}
(-\D^{(D)}-\l) v_p=\widetilde{f}_p:=f_p(1-\chi)+2\nabla u_p\cdot
\nabla\chi+u_p(\D+\l)\chi,
\end{equation*}
where $\chi=\chi(|x_1|-a)$, and in view of the established
convergence for $u_p$,
\begin{equation*}
\|v_p\|_{L_2(\Pi)}\to 1,\quad
\|\widetilde{f}_p\|_{L_2(\Pi)}\to0,\quad p\to+\infty.
\end{equation*}
It implies that $v_p$ is a singular sequence for $-\D^{(D)}$ at
$\l$.
\end{proof}

\begin{lemma}\label{lm2.2}
The discrete spectrum of $\mathcal{H}_{\mathcal{L}}$ consists of
finitely many eigenvalues.
\end{lemma}
\begin{proof}
Assume that the function $\chi$ introduced in the proof of
Lemma~\ref{lm2.1} takes the values in $[0,1]$. Due to
(\ref{2.1}) we have
\begin{equation}
\begin{aligned}
&(u,\mathcal{H}_{\mathcal{L}}u)_{L_2(\Pi)}\geqslant \|\nabla
u\|_{L_2(\Pi)}^2-c_0\|\nabla
u\|_{L_2(\Om)}^2-c_1\|u\|_{L_2(\Om)}^2\geqslant
\\
&\geqslant \big(\nabla u,(1-c_0\chi(|x_1|-a))\nabla
u\big)_{L_2(\Pi)}-\big(u,c_1\chi(|x_1|-a)u\big)_{L_2(\Pi)}.
\end{aligned}\label{2.4}
\end{equation}
We divide the tube $\Pi$ into three subsets
$\overline{\Pi}=\overline{\Pi}^{(-)}
\cup\overline{\Pi}^{(0)}\cup\overline{\Pi}^{(+)}$ defining them
as $\Pi^{(-)}:=(-\infty,-a-1)\times\om$,
$\Pi^{(+)}:=(a+1,+\infty)\times\om$,
$\Pi^{(0)}:=(-a-1,a+1)\times\om$. By
$\mathcal{H}_{\mathcal{L}}^{(\pm)}$ we indicate the Laplacian in
$L_2(\Pi^{(\pm)})$ whose domain is the subset of the functions
from $\Ho^2(\Pi^{(\pm)},\p\Pi^{(\pm)}\setminus{(\pm a\pm
1)}\times\om)$ satisfying Neumann boundary condition on $(\pm
a\pm 1)\times\om$. The symbol $\mathcal{H}_{\mathcal{L}}^{(0)}$
denotes the operator
\begin{equation*}
-\dvr\big(1-c_0\chi(|x_1|-a)\big)\nabla-c_1\chi(|x_1|-a)
\end{equation*}
in $L_2(\Pi^{(0)})$ with the domain formed by the functions from
$\Ho^2(\Pi^{(0)},(-a-1,a+1)\times\p\om)$ satisfying Neumann
condition on $\{-a-1\}\times\om$ and $\{a+1\}\times\om$. The
inequality (\ref{2.4}) implies that
\begin{equation}\label{2.5}
\mathcal{H}_{\mathcal{L}}\geqslant
\widehat{\mathcal{H}}_{\mathcal{L}}:=
\mathcal{H}_{\mathcal{L}}^{(-)}\oplus
\mathcal{H}_{\mathcal{L}}^{(0)}\oplus
\mathcal{H}_{\mathcal{L}}^{(+)}.
\end{equation}
It is easy to see that $\mathcal{H}_{\mathcal{L}}^{(\pm)}$ are
self-adjoint operators and
$\spec(\mathcal{H}_{\mathcal{L}}^{(\pm)})=
\essspec(\mathcal{H}_{\mathcal{L}}^{(\pm)})=[\nu_1,+\infty)$.
The self-adjoint operator $\mathcal{H}_{\mathcal{L}}^{(0)}$ is
lower semibounded due to (\ref{1.3}), and its spectrum is purely
discrete. Moreover, the operator $\mathcal{H}_\mathcal{L}^{(0)}$
has finitely many eigenvalues in $(-\infty,\nu_1]$. Hence, the
discrete spectrum of $\widehat{\mathcal{H}}_{\mathcal{L}}$
contains finitely many eigenvalues. Due to (\ref{2.5}) and the
minimax principle we can claim that the $k$-th eigenvalue of
$\widehat{\mathcal{H}}_{\mathcal{L}}$ is estimated from above by
the $k$-th eigenvalue of $\mathcal{H}_{\mathcal{L}}$. The former
having finitely many discrete eigenvalues, the same is true for
 $\mathcal{H}_{\mathcal{L}}$.
\end{proof}

The statement of Theorem~\ref{th1.1} follows from
Lemmas~\ref{lm2.1},~\ref{lm2.2}, if one chooses
$\mathcal{L}=\mathcal{L}_+$, $\Om=\Om_+$;
$\mathcal{L}=\mathcal{L}_-$, $\Om=\Om_-$;
$\mathcal{L}=\mathcal{L}_l$, $\Om=\om\times(-l-a_-,l+a_+)$.

\section{Analysis of $\mathcal{H}_\pm$}

In this section we establish certain properties of the operators
$\mathcal{H}_\pm$ which will be employed in the proof of
Theorems~\ref{th1.2}-\ref{th1.8}.

By $\mathbb{S}_\d$ we indicate the set of all complex numbers
separated from the half-line $[\nu_1,+\infty)$ by a distance
greater than $\d$. We choose $\d$ so that
$\discspec(\mathcal{H}_\pm)\subset\mathbb{S}_\d$.

\begin{lemma}\label{lm3.2}
The operator $(\mathcal{H}_\pm-\l)^{-1}: L_2(\Pi)\to\Ho^2(\Pi)$
is bounded and meromorphic w.r.t. $\l\in \mathbb{S}_\d$. The
poles of this operator are the eigenvalues of $\mathcal{H}_\pm$.
For any $\l$ close to $p$-multiple eigenvalue $\l_*$ of
$\mathcal{H}_\pm$ the representation
\begin{equation}\label{3.4}
(\mathcal{H}_\pm-\l)^{-1}=-\sum\limits_{j=1}^{p}
\frac{\psi_j^\pm(\cdot,\psi_j^\pm)_{L_2(\Pi)}}{\l-\l_*}+
\mathcal{T}_4^\pm(\l)
\end{equation}
holds true. Here $\psi_j^\pm$ are the eigenfunctions associated
with $\l^\pm$ and orthonormalized in $L_2(\Pi)$, while
$\mathcal{T}_4^\pm(\l): L_2(\Pi)\to\Ho^2(\Pi)$ is a bounded
operator being holomorphic w.r.t. $\l$ in a small neighbourhood
of $\l^\pm$. The relations
\begin{equation}\label{3.1a}
(\mathcal{T}_4^\pm(\l)f,\psi_j^\pm)_{L_2(\Pi)}=0,\quad
j=1,\ldots,p,
\end{equation}
are valid.
\end{lemma}

\begin{proof}
According to the results of \cite[Ch. V, Sec. 3.5]{K}, the
operator $(\mathcal{H}_\pm-\l)^{-1}$ considered as an operator
$L_2(\Pi)$ is bounded and meromorphic w.r.t.
$\l\in\mathbb{S}_\d$ and its poles are the eigenvalues of
$\mathcal{H}_\pm$. It is not difficult to check that
\begin{equation*}
(\mathcal{H}_\pm-\l-\eta)^{-1}-(\mathcal{H}_\pm-\l)^{-1}=\eta
(\mathcal{H}_\pm-\l)^{-1}(\I-\eta(\mathcal{H}_\pm-\l)^{-1})^{-1}
(\mathcal{H}_\pm-\l)^{-1}.
\end{equation*}
This identity implies that the operator
$(\mathcal{H}_\pm-\l)^{-1}$ is also bounded and meromorphic
w.r.t. $\l\in\mathbb{S}_\d$ as an operator into $\Ho^2(\Pi)$,
and its poles are the eigenvalues of $\mathcal{H}_\pm$. Consider
$\l$ ranging in a small neighbourhood of $\l_*$. The formula
(3.21) in \cite[Ch. V, Sec. 3.5]{K} gives rise to the
representation (\ref{3.4}), where the operator
$\mathcal{T}_4^\pm(\l):L_2(\Pi)\to L_2(\Pi)$ is bounded and
holomorphic w.r.t. $\l$. The self-adjointness of
$\mathcal{H}_\pm$ and (\ref{3.4}) yield that for any $f\in
L_2(\Pi)$ the identities
\begin{equation}\label{3.5}
\mathcal{T}_4^\pm(\l)f=(\mathcal{H}_\pm-\l)^{-1}\widetilde{f},\quad
\widetilde{f}=f-
\sum\limits_{j=1}^{p}\psi_j^\pm(f,\psi_j^\pm)_{L_2(\Pi)},
\end{equation}
and (\ref{3.1a}) hold true. Let $\Psi^\perp$ be a subspace of
the functions in $L_2(\Pi)$ which are orthogonal to
$\psi_j^\pm$, $j=m,\ldots,m+p-1$. The identities (\ref{3.5})
mean that
\begin{equation}\label{3.2c}
\mathcal{T}_4^\pm(\l)\big|_{\Psi^\perp}=
(\mathcal{H}_\pm-\l)^{-1}\big|_{\Psi^\perp}.
\end{equation}
The operator $(\mathcal{H}_\pm-\l)^{-1}\big|_{\Psi^\perp}$ is
holomorphic w.r.t. $\l$ as an operator from $\Psi^\perp$ into
$\Ho^2(\Pi)\cap \Psi^\perp$. This fact is due to holomorphy in
$\l$ of the operator $(\mathcal{H}_\pm-\l):\Ho^2(\Pi)\cap
\Psi^\perp\to \Psi^\perp$ and the invertibility of this operator
(see \cite[Ch. V\!I\!I, Sec. 1.1]{K}). Therefore, the
restriction of $\mathcal{T}_4^\pm(\l)$ on $\Psi^\perp$ is
holomorphic w.r.t. $\l$ as an operator into $\Ho^2(\Pi)$. Taking
into account (\ref{3.5}), we conclude that for any $f\in
L_2(\Pi)$ the function $\mathcal{T}_4^\pm(\l)f$ is holomorphic
w.r.t. $\l$. The holomorphy in a weak sense implies the
holomorphy in the norm sense \cite[Ch. V\!I\!I, Sec. 1.1]{K},
and we arrive at the statement of the lemma.
\end{proof}

Let $0<\nu_1<\nu_2\leqslant\ldots\leqslant\nu_j\leqslant\ldots$
be the eigenvalues of the negative Dirichlet Laplacian in $\om$
taken in a non-decreasing order counting multiplicity, and
$\phi_i=\phi_i(x')$ be the associated eigenfunctions
orthonormalized in $L_2(\om)$. We denote $\Pi_a^\pm:=\Pi\cap\{x:
\pm x_1>\pm a\}$.

\begin{lemma}\label{lm3.1}
Suppose that $u\in\H^1(\Pi_a^\pm)$ is a solution to the boundary
value problem
\begin{equation*}
(\D+\l)u=0,\quad x\in\Pi_a^\pm,\qquad u=0,\quad
x\in\p\Pi\cap\overline{\Pi}_a^\pm,
\end{equation*}
where $\l\in\mathbb{S}_\d$. Then the function $u$ can be
represented as
\begin{equation}
u(x)=\sum\limits_{j=1}^{\infty} \a_j\E^{-s_j(\l)(\pm
x_1-a)}\phi_j(x'), \quad \a_j=\int\limits_\om
u(a,x')\phi_j(x')\di x'. \label{3.1}
\end{equation}
The series (\ref{3.1}) converges in $\H^p(\Pi_b^\pm)$ for any
$\Pi_b^\pm\subset\Pi_a^\pm$, $p\geqslant 0$. The coefficients
$\a_j$ satisfy the identity
\begin{equation}\label{3.3}
\sum\limits_{j=1}^{\infty}|\a_j|^2=\|u(\cdot,a)\|_{L_2(\om)}^2.
\end{equation}
\end{lemma}

\begin{proof}
In view of the obvious change of variables it is sufficient to
prove the lemma for $\Pi_0^+$. It is clear that $v\in
C^\infty(\overline{\Pi}_0^+\setminus\om_0)$, where
$\om_0:=\{0\}\times\overline{\om}$. Since $v\in\H^1(\Pi_0^+)$
and $v(x,\l)=0$ as $x\in\p\Pi_0^+\setminus\om_0$, the
representation
\begin{equation*}
u(x,\l)=\sum\limits_{j=1}^{\infty}\phi_j(x')\int\limits_\om
u(x_1,t,\l)\phi_j(t)\di t
\end{equation*}
holds true for any $x_1\geqslant 0$ in $L_2(\om)$. We have the
identity
\begin{equation}\label{3.31}
\sum\limits_{j=1}^{\infty}\bigg| \int\limits_\om
u(x,\l)\phi_j(x')\di x'
\bigg|^2=\|u(x_1,\cdot,\l)\|_{L_2(\om)}^2
\end{equation}
for any $x_1\geqslant 0$. Employing the equation for $u$ we
obtain
\begin{align*}
&\frac{d^2}{dx_1^2}\int\limits_\om u(x,\l)\phi_j(x')\di x'
=-\int\limits_\om\phi_j(x')\left(\D_{x'}+\l\right) u(x,\l)\di x'
\\
&=-\int\limits_\om u(x,\l)\left(\D_{x'}+\l\right)\phi_j(x')\di
x'=(\nu_j-\l)\int\limits_\om u(x,\l)\phi_j(x')\di x',\quad
x_1>0.
\end{align*}
The function $v$ belonging to $\H^1(\Pi_0^+)$, the relation
obtained implies that
\begin{equation}\label{3.9a}
\int\limits_\om u(x,\l)\phi_j(x') \di x'=
\a_{j}\E^{-\sqrt{\nu_j-\l}x_1}.
\end{equation}
We have employed here the identity
\begin{equation*}
\lim\limits_{x_1\to +0}\int\limits_0^\pi u(x,\l)\phi_j(x')\di
x'=\int\limits_\om u(0,x',\l)\phi_j(x') \di x',
\end{equation*}
which follows from the estimate
\begin{align*}
&\left|\int\limits_\om \big(u(x,\l)-u(0,x',\l)\big)\phi_j(x')\di
x'\right|
\\
&\hphantom{\int\limits_\om \big(u(x,\l)}
\leqslant\left|\int\limits_\om \int\limits_0^{x_1} \frac{\p
u}{\p t_1}(t,x')\phi_j(x')\di t_1\di x'\right| \leqslant
\sqrt{|x_1|}\left\|\frac{\p u}{\p x_1}\right\|_{L_2(\Pi_0^+)}.
\end{align*}
The identities (\ref{3.31}), (\ref{3.9a}) yield (\ref{3.3}). For
$x_1\geqslant b>0$ the coefficients of  (\ref{3.1}) decays
exponentially as $j\to+\infty$, that implies the convergence of
the series in (\ref{3.1}) in $\H^p(\Pi_b^+)$, $p\geqslant 0$.
\end{proof}

For $l\geqslant a_-+a_+$ we introduce the operators
$\mathcal{T}_6^\pm(\l,l): L_2(\Pi,\Om_\mp)\to L_2(\Pi,\Om_\pm)$,
\begin{equation}\label{3.6a}
\mathcal{T}_6^\pm(\l,l):=\mathcal{L}_\pm \mathcal{S}(\pm 2l)
(\mathcal{H}_\mp-\l)^{-1}.
\end{equation}

\begin{lemma}\label{lm3.4}
The operator $\mathcal{T}_6^\pm$ is bounded and meromorphic
w.r.t. $\l\in\mathbb{S}_\d$. For any compact set
$\mathbb{K}\subset \mathbb{S}_\d$ separated from
$\discspec(\mathcal{H}_\mp)$ by a positive distance the
estimates
\begin{equation}\label{3.7}
\left\|\frac{\p^i \mathcal{T}_6^\pm}{\p\l^i}\right\|\leqslant
Cl^i\E^{-2 l \RE s_1(\l)},\quad i=0,1, \l\in\mathbb{K},
\end{equation}
hold true, where the constant $C$ is independent of $\l\in
\mathbb{K}$ and $l\geqslant a_-+a_+$.  For any $\l$ close to a
$p$-multiple eigenvalue $\l_*$ of  $\mathcal{H}_\mp$ the
representation
\begin{equation}
\mathcal{T}_6^\pm(\l,l)=-\sum\limits_{j=1}^{p}
\frac{\vp_j^\mp(\cdot,\psi_j^\mp)_{L_2(\Om_\mp)}}{\l-\l_*}+
\mathcal{T}_7^\pm(\l,l),\quad \vp_j^\mp:=\mathcal{L}_\pm
\mathcal{S}(\pm 2l)\psi_j^\mp,\label{3.8a}
\end{equation}
is valid. Here $\psi_j^\mp$ are the eigenfunctions associated
with $\l_*$ and orthonormalized in $L_2(\Pi)$, while the
operator $\mathcal{T}_7^\pm(\l,l): L_2(\Pi,\Om_\mp)\to
L_2(\Pi,\Om_\pm)$ is bounded and holomorphic w.r.t. $\l$ close
to $\l_*$ and satisfies the estimates
\begin{equation}\label{3.9}
\left\|\frac{\p^i \mathcal{T}_7^\pm}{\p\l^i}\right\|\leqslant
Cl^{i+1}\E^{-2l\RE s_1(\l)},\quad i=0,1,
\end{equation}
where the constant $C$ is independent of $\l$ close to $\l_*$
and $l\geqslant a_-+a_+$. The identities
\begin{equation}\label{3.10}
\mathcal{T}_7^\pm(\l_*,l)=\mathcal{L}_\pm \mathcal{S}(\pm
2l)\mathcal{T}_4^\mp(\l_*)
\end{equation}
hold true.
\end{lemma}

\begin{proof}
We prove the lemma for $\mathcal{T}_6^+$ only; the proof for
$\mathcal{T}_6^-$ is similar. Let $f\in L_2(\Pi,\Om_-)$, and
denote $u:=(\mathcal{H}_--\l)^{-1}f$. The function $f$ having
compact support, by Lemma~\ref{lm3.1} the function $u$ can be
represented as the series (\ref{3.1}) for $x_1\geqslant a_-$.
Hence,
\begin{equation*}
\big(\mathcal{S}(2l) u\big)(x)= \sum\limits_{j=1}^{\infty} \a_j
\E^{-2s_j(\l) l} \E^{-s_j(\l)(\pm x_1-a)}\phi_j(x').
\end{equation*}
Employing this representation and (\ref{3.3}), we obtain
\begin{align}
&\left\|\sum\limits_{j=1}^{\infty}
\a_j\E^{-2s_j(\l)l}\E^{-s_j(\l)(x_1-
a_-)}\phi_j(x')\right\|_{\H^2(\Om_+)} \leqslant\nonumber
\\
&\leqslant C\sum\limits_{j=1}^{\infty} |\a_j| \E^{-2l\RE
s_j(\l)} \|\E^{-s_j(\l)(\cdot-a_-)}\|_{\H^2(-a_+,a_+)}
\left(\|\D_{x'}\phi_j\|_{L_2(\om)}+\|\phi_j\|_{L_2(\om)}\right)
\leqslant\nonumber
\\
&\leqslant C\sum\limits_{j=1}^{\infty}
|\a_j||s_j(\l)|^{3/2}\nu_j\E^{-(2l-a_--a_+)\RE
s_j(\l)}\leqslant\nonumber
\\
&\leqslant C\left(\sum\limits_{j=1}^{\infty}
|\a_j|^2\right)^{1/2}
\left(\sum\limits_{j=1}^{\infty}(\nu_j^{7/2}+|\l|^{7/2})
\E^{-2(2l-a_--a_+)\RE s_j(\l)}\right)^{1/2}\leqslant \nonumber
\\
&\leqslant C(1+|\l|^{7/4})\E^{-(2l-a_--a_+)\RE s_1(\l)}
\|u(a,\cdot)\|_{L_2(\om)},\label{3.6}
\end{align}
where the constant $C$ is independent of $\l\in \mathbb{S}_\d$
and $l\geqslant  a_-+a_+$. Here we have also applied the
well-known estimate \cite[Ch. I\!V, Sec. 1.5, Theorem 5]{M}
\begin{equation}\label{3.6b}
cj^{\frac{2}{n-1}}\leqslant \nu_j\leqslant Cj^{\frac{2}{n-1}}.
\end{equation}
By direct calculations we check that
\begin{equation*}
\left(\frac{\p \mathcal{S}(2l) u}{\p\l}\right)(x,\l,l)=
\sum\limits_{j=1}^{\infty}\frac{\a_j}{2s_j(\l)}(x_1-a_-+2l)
\E^{-2s_j(\l)l}\E^{-s_j(\l)(x_1-a_-)}\phi_j(x').
\end{equation*}
Proceeding in the same way as in (\ref{3.6}) we obtain
\begin{equation}\label{3.7a}
\left\|\frac{\p \mathcal{S}(2l) u}{\p\l}\right\|_{\H^2(\Om_+)}
\leqslant C l(1+|\l|^{5/4})\E^{-(2l-a_--a_+)\RE
s_1(\l)}\|u(a,\cdot)\|_{L_2(\om)},
\end{equation}
where the constant $C$ is independent of $\l\in \mathbb{S}_\d$
and $l\geqslant  a_-+a_+$. In the same way we check that
\begin{equation}\label{3.7b}
\left\|\frac{\p^2\mathcal{S}(2l)
u}{\p\l^2}\right\|_{\H^2(\Om_+)} \leqslant C
l(1+|\l|^{3/4})\E^{-(2l-a_--a_+)\RE
s_1(\l)}\|u(a,\cdot)\|_{L_2(\om)},
\end{equation}
where the constant $C$ is independent of $\l\in \mathbb{S}_\d$
and $l\geqslant  a_-+a_+$. Lemma~\ref{lm3.2} implies
\begin{equation*}
u(a_-,\cdot)=-\sum\limits_{j=1}^{p}
\frac{\psi_j^-(a_-,\cdot)(f,\psi_j^-)_{L_2(\Pi)}}{\l-\l_*}+
\big(\mathcal{T}_4^-(\l)f\big)(a_-,\cdot).
\end{equation*}
This representation and (\ref{3.3}), (\ref{3.6}), (\ref{3.7a}),
(\ref{3.7b}) lead us to the statement of the lemma.
\end{proof}

\section{Reduction of the eigenvalue equation
for  $\mathcal{H}_l$}

In this section we reduce the eigenvalue equation
\begin{equation}\label{4.1}
\mathcal{H}_l\psi=\l\psi
\end{equation}
to an operator equation in the space $L_2(\Om_-)\oplus
L_2(\Om_+)$. The reduction will be one of the key ingredients in
the proofs of Theorems~\ref{th1.2}-\ref{th1.8}. Hereafter we
assume that $l\geqslant a_-+a_+$.

Let $f_\pm\in L_2(\Pi,\Om_\pm)$ be a pair of arbitrary
functions, and let the functions $u_\pm$ satisfy the equations
\begin{equation}\label{4.2}
(\mathcal{H}_\pm-\l)u_\pm=f_\pm,
\end{equation}
where $\l\in\mathbb{S}_\d$. We choose $\d$ so that
$\discspec(\mathcal{H}_\pm)\subset \mathbb{S}_\d$. We construct
a solution to (\ref{4.1}) as
\begin{equation}\label{4.3}
\psi=\mathcal{S}(l)u_-+\mathcal{S}(-l)u_+.
\end{equation}
Suppose that the function $\psi$ defined in this way satisfies
(\ref{4.1}). We substitute (\ref{4.3}) into (\ref{4.1}) to
obtain
\begin{equation*}
0=(\mathcal{H}_l-\l)\psi=(\mathcal{H}_l-\l)\mathcal{S}(l)u_-+
(\mathcal{H}_l-\l)\mathcal{S}(-l)u_+.
\end{equation*}
By direct calculations we check that
\begin{align*}
(\mathcal{H}_l-\l)\mathcal{S}(l)u_-&=
\mathcal{S}(l)(-\D^{(D)}+\mathcal{L}_--\l)\mathcal{S}(-l)
\mathcal{S}(l)u_-
+\mathcal{S}(-l)\mathcal{L}_+\mathcal{S}(2l)u_-=
\\
&=
\mathcal{S}(l)f_-+\mathcal{S}(-l)\mathcal{L}_+\mathcal{S}(2l)u_-,
\\
(\mathcal{H}_l-\l)\mathcal{S}(-l)u_+&=\mathcal{S}(-l)f_+ +
\mathcal{S}(l)\mathcal{L}_-\mathcal{S}(-2l)u_+.
\end{align*}

Hence,
\begin{equation*}
\mathcal{S}(l)\big(f_- + \mathcal{L}_-\mathcal{S}(-2l)u_+\big)+
\mathcal{S}(-l)\big(f_+ +\mathcal{L}_+\mathcal{S}(2l)u_-\big)=0.
\end{equation*}
Since the functions $f_\pm+\mathcal{L}_\pm\mathcal{S}(\pm
2l)u_\mp$ are compactly supported, it follows that the functions
$\mathcal{S}(\mp l)\big(f_\pm+ \mathcal{L}_\pm\mathcal{S}(\pm
2l)u_\mp\big)$ are compactly supported, too, and their supports
are disjoint. Thus, the last equation obtained is equivalent to
the pair of the equations
\begin{equation}\label{4.4}
f_- + \mathcal{L}_-\mathcal{S}(-2l)u_+=0, \quad f_+ +
\mathcal{L}_+\mathcal{S}(2l)u_-=0.
\end{equation}
These equations are equivalent to the equation (\ref{4.1}). The
proof of this fact is the subject of

\begin{lemma}\label{lm4.2}
To any solution $\boldsymbol{f}:=(f_-,f_+)\in L_2(\Om_-)\oplus
L_2(\Om_+)$ of (\ref{4.4}) and functions $u_\pm$ solving
(\ref{4.2}) there exists a unique solution of (\ref{4.1}) given
by (\ref{4.2}), (\ref{4.3}). For any solution $\psi$ of
(\ref{4.1}) there exists a unique $\boldsymbol{f}\in
L_2(\Om_-)\oplus L_2(\Om_+)$ solving (\ref{4.4}) and the unique
functions $u_\pm$ satisfying (\ref{4.2}) such that $\psi$ is
given by (\ref{4.2}), (\ref{4.3}). The equivalence holds for any
$\l\in \mathbb{S}_\d$.
\end{lemma}

\begin{proof}
It was shown above that if $\boldsymbol{f}\in L_2(\Om_-)\oplus
L_2(\Om_+)$ solves (\ref{4.4}) and the functions $u_\pm$ are the
solutions to (\ref{4.2}), the function $\psi$ defined by
(\ref{4.3}) solves (\ref{4.1}).

Suppose that $\psi$ is a solution of (\ref{4.1}). This functions
satisfies the equation
\begin{equation}\label{4.6}
(-\D-\l)\psi=0,\qquad -l+a_-<x_1<l-a_+,\quad x'\in\om,
\end{equation}
and vanishes as $-l+a_-<x_1<l-a_+$, $x'\in\p\om$. Due to
standard smoothness improving theorems (see, for instance,
\cite[Ch. I\!V, Sec. 2.2]{M}) it implies that $\psi\in
C^\infty(\{x: -l+a_-<x_1<l-a_+,\quad x'\in\overline{\om}\})$.
Hence, the numbers
\begin{equation*}
\rho_j^\pm=\rho_j^\pm(l):=\frac{1}{2}\int\limits_\om
\left(\psi(0,x',l)\pm \frac{1}{s_j(\l)}\frac{\p\psi}{\p
x_1}(0,x',l)\right)\phi_j(x')\di x'
\end{equation*}
are well-defined. Employing the equation (\ref{4.6}), the
identity $\psi=0$ as $x\in\p\Pi$, and the smoothness of $\psi$
we integrate by parts,
\begin{align*}
\rho_j^\pm&=-\frac{1}{2\nu_j}\int\limits_\om
\left(\psi(0,x',l)\pm \frac{1}{s_j(\l)}\frac{\p\psi}{\p
x_1}(0,x',l)\right)\D_{x'}\phi_j(x')\di x'=
\\
&= -\frac{1}{2\nu_j} \int\limits_\om
\phi_j(x')\D_{x'}\left(\psi(0,x',l)\pm
\frac{1}{s_j(\l)}\frac{\p\psi}{\p x_1}(0,x',l)\right)\di x'=
\\
&=\frac{1}{2\nu_j} \int\limits_\om
\phi_j(x')\left(\frac{\p^2}{\p x_1^2}+\l\right)\left(\psi\pm
\frac{1}{s_j(\l)}\frac{\p\psi}{\p x_1}\right)\Bigg|_{x_1=0}\di
x'=
\\
&=\frac{1}{2\nu_j^p} \int\limits_\om
\phi_j(x')\left(\frac{\p^2}{\p x_1^2}+\l\right)^p\left(\psi\pm
\frac{1}{s_j(\l)}\frac{\p\psi}{\p x_1}\right)\Bigg|_{x_1=0}\di
x'
\end{align*}
for any $p\in\mathbb{N}$. In view of (\ref{3.6b}) it yields that
$\sum\limits_{j=1}^{\infty} j^p|\rho_j^\pm|<\infty$ for any
$p\in \mathbb{N}$. Now we introduce the functions $u_\pm$,
\begin{align*}
&u_\pm(x_1\mp l,x',l):=\sum\limits_{j=1}^{\infty}
\rho_j^\pm(l)\E^{\pm s_j(\l) x_1}\phi_j(x'), && x\in\Pi_0^\mp,
\\
&u_\pm(x_1\mp l,x',l):=\psi(x,l)-u_\mp(x_1\pm l,x',l),&&
x\in\Pi_0^\pm,
\end{align*}
and conclude that
\begin{equation*}
u_-(x_1+l,x',l),u_+(x_1-l,x',l)\in
C^\infty(\overline{\Pi}_0^\pm)\cap
\Ho^2(\Pi_0^\pm,\p\Pi\cap\p\Pi_0^\pm).
\end{equation*}
The smoothness of $\psi$ gives rise to the representations
\begin{align*}
&\psi(0,x',l)=\sum\limits_{j=1}^{\infty}
\big(\rho_j^+(l)+\rho_j^-(l)\big)\phi_j(x'),
\\
&\frac{\p\psi}{\p x_1}(0,x',l)=\sum\limits_{j=1}^{\infty}
\big(\rho_j^+(l)-\rho_j^-(l)\big)s_j(\l)\phi_j(x').
\end{align*}
These relations and the aforementioned smoothness of $u_\pm$
imply that $u_\pm\in\Ho^2(\Pi)$. We also infer that the
introduced functions $u_\pm$ satisfy (\ref{4.3}). We define now
the vector $\boldsymbol{f}=(f_-,f_+)\in L_2(\Om_-)\oplus
L_2(\Om_+)$ by $f_-:=-\mathcal{L}_-u_+(x_1-2l,x',l)$,
$f_+:=-\mathcal{L}_+u_-(x_1+2l,x',l)$. Let us check that the
functions $u_\pm$ satisfy (\ref{4.2}); it will imply that the
vector $\boldsymbol{f}$ just introduced solves (\ref{4.5}).

The definition of $u_\pm$ implies that
\begin{equation*}
(\D+\l)u_\pm(x_1\mp l,x',l)=0,\quad x\in\overline{\Pi}_0^\mp.
\end{equation*}
Thus,
\begin{equation*}
(\D+\l)u_-=0,\quad x\in\overline{\Pi}_l^+,\qquad
(\D+\l)u_+=0,\quad x\in\overline{\Pi}_{-l}^-.
\end{equation*}
By these equations, the equation for $\psi$, and the definitions
of $u_-(x_1+l,x',l)$ for $x\in\Pi_0^-$, we obtain that for
$x\in\Pi_{l}^-$
\begin{align*}
&(-\D-\l+\mathcal{L}_-)u_-(x,l)=(-\D-\l+\mathcal{L}_-)
\big(\psi(x_1-l,x',l)-u_+(x_1-2l,x',l)\big)=
\\
&=-(-\D-\l+\mathcal{L}_-)u_+(x_1-2l,x',l)=-\mathcal{L}_-
u_+(x_1-2l,x',l)=f_-(x).
\end{align*}
Therefore, $(\mathcal{H}_--\l)u_-=f_-$. The relation
$(\mathcal{H}_+-\l)u_+=f_+$ can be established in the same way.
\end{proof}

Suppose that $\l\in\mathbb{S}_\d\setminus\si_*$. In this case
$u_\pm=(\mathcal{H}_\pm-\l)^{-1}f_\pm$. This fact together with
the definition (\ref{3.6a}) of $\mathcal{T}_6^\pm$ implies that
$\mathcal{L}_\pm\mathcal{S}(\pm
2l)u_\mp=\mathcal{T}_6^\pm(\l,l)f_\pm$. Substituting the
identity obtained into (\ref{4.4}), we arrive at the equation
\begin{equation}\label{4.5}
\boldsymbol{f}+\mathcal{T}_2(\l,l)\boldsymbol{f}=0,
\end{equation}
where $\boldsymbol{f}:=(f_-,f_+)\in L_2(\Om_-)\oplus
L_2(\Om_+)$, where the operator $\mathcal{T}_2:L_2(\Om_-)\oplus
L_2(\Om_+)\to L_2(\Om_-)\oplus L_2(\Om_+)$ is defined in
(\ref{1.25}).

\begin{proof}[Proof of Theorem~\ref{th1.2}]
The inequality (\ref{2.4}) yields that the operator
$\mathcal{H}_l$ is lower semibounded with lower bound $-c_1$.
Together with Theorem~\ref{th1.1} it implies that the discrete
eigenvalues of the operator are located in $[-c_1,\nu_1)$. We
introduce the set
$\mathbb{K}_\d:=[-c_1,\nu_1-\d)\setminus\bigcup\limits_{\l\in\si_*}
(\l-\d,\l+\d)$. It satisfies the hypothesis of
Lemma~\ref{lm3.4}, and the estimate (\ref{3.7}) implies that
\begin{equation*}
\|\mathcal{T}_2(\l,l)\|\leqslant C(\d)\E^{-2l\RE s_1(\l)},
\end{equation*}
if $l$ is large enough and $\l\in\mathbb{K}_\d$. The constant
$C$ in this estimate is independent of $\l\in\mathbb{K}_\d$ and
$l$ large enough. Hence, for such $\l$ and $l$ the operator
$(\I+\mathcal{T}_2(\l,l))$ is boundedly invertible. Therefore,
the equation (\ref{4.5}) has no nontrivial solutions, if $l$ is
large enough and $\l\in\mathbb{K}_\d$. Since
$\mathbb{K}_\d\cap\si_*=\emptyset$, the identity
$\boldsymbol{f}=0$ implies that
$u_\pm=(\mathcal{H}_\pm-\l)^{-1}f_\pm=0$, i.e., $\psi=0$. Thus,
the equation (\ref{4.1}) has no nontrivial solution for
$\l\in\mathbb{K}_\d$, if $l$ is large enough. Therefore,
$\mathbb{K}_\d\cap\discspec(\mathcal{H}_l)=\emptyset$, if $l$ is
large enough. The number $\d$ being arbitrary, the last identity
completes the proof.
\end{proof}

Assume that $\l_*\in\si_*$ is $(p_-+p_+)$-multiple.
Lemma~\ref{lm3.4} implies that for $\l$ close to $\l_*$ the
representation (\ref{4.11}) holds true, where the operator
$\mathcal{T}_3$ is given by
\begin{align*}
&
\mathcal{T}_3(\l,l)\boldsymbol{f}:=(\mathcal{T}_7^-(\l,l)f_+,
\mathcal{T}_6^+(\l,l)f_-),\quad\text{if}\quad p_-=0,
\\
&\mathcal{T}_3(\l,l)\boldsymbol{f}:=(\mathcal{T}_6^-(\l,l)f_+,
\mathcal{T}_7^+(\l,l)f_-),\quad\text{if}\quad p_+=0,
\\
& \mathcal{T}_3(\l,l)\boldsymbol{f}:=(\mathcal{T}_7^-(\l,l)f_+,
\mathcal{T}_7^+(\l,l)f_-),\quad\text{if}\quad p_-\not=0,\quad
p_+\not=0.
\end{align*}
Lemma~\ref{lm3.4} yields also that the operator
$\mathcal{T}_3(\l,l)$ is bounded and holomorphic w.r.t. $\l$ in
a small neighbourhood of $\l_*$, and satisfies the estimate
\begin{equation}\label{4.12}
\left\|\frac{\p^i \mathcal{T}_3}{\p\l^i}\right\|\leqslant
Cl^{i+1}\E^{-2l\RE s_1(\l)},\quad i=0,1,
\end{equation}
where the constant $C$ is independent of $l\geqslant a_-+a_+$
and $\l$ close to $\l_*$.

Suppose that $\l\not=\l_*$ is an eigenvalue of $\mathcal{H}_l$
converging to $\l_*$. In this case the identity
$\boldsymbol{f}=0$ leads us to the relations
$u_\pm=(\mathcal{H}_\pm-\l)^{-1}f_\pm=0$, $\psi=0$. Thus, the
corresponding equation (\ref{4.5}) has a nontrivial solution.
Let us solve this equation.

We substitute (\ref{4.11}) into (\ref{4.5}), and arrive at the
following equation:
\begin{equation}\label{4.8a}
\boldsymbol{f}-\frac{1}{\l-\l_*}\sum\limits_{i=1}^{p}
\boldsymbol{\phi}_i
\mathcal{T}_1^{(i)}\boldsymbol{f}+\mathcal{T}_3\boldsymbol{f}=0.
\end{equation}
In view of the estimate (\ref{4.12}) the operator
$(\I+\mathcal{T}_3(\l,l))$ is invertible, and the operator
$(\I+\mathcal{T}_3)^{-1}$ is bounded and holomorphic w.r.t. $\l$
in a small neighbourhood of $\l_*$. Applying this operator to
the last equation gives rise to one more equation,
\begin{equation}\label{4.13}
\boldsymbol{f}=\frac{1}{\l-\l_*}\sum\limits_{i=1}^{p}
\boldsymbol{\Phi}_i \mathcal{T}_1^{(i)}\boldsymbol{f},
\end{equation}
where $\boldsymbol{\Phi}_i=\boldsymbol{\Phi}_i(\cdot,\l,l):=
(\I+\mathcal{T}_3(\l,l))^{-1}\boldsymbol{\phi}_i(\cdot,l)$. We
denote
$k_i=k_i(\l,l):=(\l-\l_*)^{-1}\mathcal{T}_1^{(i)}\boldsymbol{f}$,
and apply the functionals $\mathcal{T}_1^{(j)}$, $j=1,\ldots,p$,
to the equation (\ref{4.13}) that leads us to a system of linear
equations (\ref{4.14}), where $\boldsymbol{k}:=(k_1\ldots
k_p)^{t}$.

Given a non-trivial solution of (\ref{4.5}), the associated
vector $\boldsymbol{k}$ is non-zero, since otherwise the
definition of $k_i$, and (\ref{4.13}) would imply that
$\boldsymbol{f}=0$. Therefore, if $\l\not=\l_*$ is an eigenvalue
of $\mathcal{H}_l$, the system (\ref{4.14}) has a non-trivial
solution. It is true, if and only if the equation (\ref{4.15})
holds true. Thus, each eigenvalue of $\mathcal{H}_l$ converging
to $\l_*$ and not coinciding with $\l_*$ should satisfy this
equation.

Let us show that if $\l_*$ is an eigenvalue of $\mathcal{H}_l$,
it satisfies (\ref{4.15}) as well. In this case $\l_*$ the
associated eigenfunction is given by (\ref{4.3}) that is due to
Lemma~\ref{lm4.2}. The self-adjointness of $\mathcal{H}_\pm$
implies that
\begin{equation}\label{4.15a}
(f_\pm,\psi_i^\pm)_{L_2(\Om_\pm)}=\big((\mathcal{H}_\pm-\l_*)u_\pm,
\psi_i^\pm\big)_{L_2(\Om_\pm)}=
\big(u_\pm,(\mathcal{H}_\pm-\l_*)\psi_i^\pm\big)_{L_2(\Om_\pm)}=0,
\end{equation}
$i=1,\ldots,p_\pm$. Therefore, the functions $u_\pm$ can be
represented as
\begin{equation}\label{4.16}
u_-(\cdot,l)=\mathcal{T}_4^-(\l_*)f_--\sum\limits_{i=1}^{p_-}k_i
\psi_i^-,\quad
u_+(\cdot,l)=\mathcal{T}_4^+(\l_*)f_+-\sum\limits_{i=1}^{p_+}k_{i+p_-}
\psi_i^+,
\end{equation}
where $k_i$ are numbers to be found. Employing the relation
(\ref{3.10}) and substituting (\ref{4.16}) into (\ref{4.4}), we
obtain the equation
\begin{equation*}
\boldsymbol{f}+\mathcal{T}_3(\l_*,l)\boldsymbol{f}
=\sum\limits_{i=1}^{p} k_i\boldsymbol{\phi}_i(\cdot,l),
\end{equation*}
which is equivalent to
\begin{equation}\label{4.17}
\boldsymbol{f}=\sum\limits_{i=1}^{p}k_i
\boldsymbol{\Phi}_i(\cdot,\l_*,l).
\end{equation}
The relations (\ref{4.15a}) can be rewritten as
$\mathcal{T}_1^{(i)}\boldsymbol{f}=0$, $i=1,\ldots,p$, that
together with (\ref{4.17}) implies the system (\ref{4.14}) for
$\l=\l_*$. The vector $\boldsymbol{k}$ is non-zero since
otherwise the relations (\ref{4.17}), (\ref{4.16}) would imply
$\boldsymbol{f}=0$, $u_\pm=0$, $\psi=0$. Thus,
$\det\mathrm{A}(\l_*,l)=0$, which coincides with the equation
(\ref{4.15}) for $\l=\l_*$.

Let $\l$ be a root of (\ref{4.15}), converging to $\l_*$ as
$l\to+\infty$. We are going to prove that in this case the
equation (\ref{4.1}) has a non-trivial solution, i.e., $\l$ is
an eigenvalue of $\mathcal{H}_l$. The equation  (\ref{4.1})
being satisfied, it follows that the system (\ref{4.14}) has a
nontrivial solution $\boldsymbol{k}$. We specify this solution
by the requirement
\begin{equation}\label{4.19}
\|\boldsymbol{k}\|_{\mathbb{C}^p}=1,
\end{equation}
and define $\boldsymbol{f}:=\sum\limits_{i=1}^{p}k_i
\boldsymbol{\Phi}_i(\cdot,\l,l)\in\boldsymbol{f}\in
L_2(\Om_-)\oplus L_2(\Om_+)$. The system (\ref{4.14}) and the
definition of $\mathcal{T}_1^{(i)}$ give rise to the identities
\begin{equation}\label{4.19a}
\begin{aligned}
&(f_-,\psi_i^-)_{L_2(\Om_-)}=\mathcal{T}_1^{(i)}\boldsymbol{f}
=(\l-\l_*)k_i,&& i=1,\ldots,p_-,
\\
&(f_+,\psi_i^+)_{L_2(\Om_-)}=\mathcal{T}_1^{(i+p_-)}\boldsymbol{f}
=(\l-\l_*)k_{i+p_-},&& i=1,\ldots,p_+.
\end{aligned}
\end{equation}
Taking these identities into account and employing (\ref{3.4}),
in the case $\l\not=\l_*$ we arrive at the formulas
\begin{equation}\label{4.20}
u_-=-\sum\limits_{i=1}^{p_-} k_i\psi_i^- +
\mathcal{T}_4^-(\l)f_-, \quad u_+=-\sum\limits_{i=1}^{p_+}
k_{i+p_-}\psi_i^+ + \mathcal{T}_4^+(\l)f_+.
\end{equation}
In the case $\l=\l_*$ we adopt these formulas as the definition
of the functions $u_\pm$ that is possible due to the identities
(\ref{4.19a}) with $\l=\l_*$ and (\ref{3.2c}).

If $\l\not=\l_*$, we employ the system (\ref{4.14}) to check by
direct calculations that $\boldsymbol{f}$ solves (\ref{4.13}),
and thus the equations (\ref{4.4}). If $\l=\l_*$, the identity
(\ref{3.10}) implies
\begin{align*}
\big(\mathcal{L}_-&\mathcal{S}(-2l)u_+,
\mathcal{L}_+\mathcal{S}(2l)u_-\big)=\mathcal{T}_3(\l_*,l)
\boldsymbol{f} -
\sum\limits_{i=1}^{p}k_i\boldsymbol{\phi}_i(\cdot,l)=
\mathcal{T}_3(\l_*,l)\boldsymbol{f}
\\
&-\sum\limits_{i=1}^{p}k_i
(\I+\mathcal{T}_3(\l_*,l))\boldsymbol{\Phi}_i(\cdot,\l_*,l)=
\mathcal{T}_3(\l_*,l)\boldsymbol{f}-(\I+\mathcal{T}_3(\l_*,l))
\boldsymbol{f} = -\boldsymbol{f}.
\end{align*}
Hence, the equation (\ref{4.4}) holds true. With
Lemma~\ref{lm4.2} in mind we therefore conclude that the
function $\psi$ defined by (\ref{4.3}) solves (\ref{4.1}).

Let us prove that $\psi\not\equiv0$; it will imply that $\l$ is
an eigenvalue of $\mathcal{H}_l$. Lemma~\ref{lm1.1} implies that
the functions $\vp_i^\pm$ obey the estimate
\begin{equation}\label{5.1}
\|\vp_i^\pm\|_{L_2(\Om_\mp)}\leqslant C\E^{-2ls_1(\l_*)},
\end{equation}
where the constant $C$ is independent of $l$. This estimate
together with (\ref{4.12}) gives rise to the similar estimates
for $\boldsymbol{\Phi}_i$:
\begin{equation}\label{5.2}
\begin{aligned}
&\|\boldsymbol{\Phi}_i\|_{L_2(\Om_-)\oplus L_2(\Om_+)}\leqslant
C\E^{-2ls_1(\l_*)},
\\
&\left\|\frac{\p\boldsymbol{\Phi}_i}{\p\l}
(\cdot,\l,l)\right\|_{L_2(\Om_-)\oplus L_2(\Om_+)}\leqslant
Cl^2\E^{-2l(s_1(\l_*)+s_1(\l))},
\end{aligned}
\end{equation}
where the constant $C$ is independent of $\l$ and $l$. The
latter inequality is based on the formula
\begin{equation*}
\frac{\p}{\p\l}(\I+\mathcal{T}_3)^{-1}=- (\I+\mathcal{T}_3)^{-1}
\frac{\p \mathcal{T}_3}{\p\l} (\I+\mathcal{T}_3)^{-1},
\end{equation*}
which follows from the obvious identities:
\begin{align*}
\I=(\I+\mathcal{T}_3)(\I+\mathcal{T}_3)^{-1},\quad
0=\frac{\p\mathcal{T}_3}{\p\l}(\I+\mathcal{T}_3)^{-1}+
(\I+\mathcal{T}_3)\frac{\p}{\p\l}(\I+\mathcal{T}_3)^{-1}.
\end{align*}
Hence, the asymptotics (\ref{1.8}) is valid. The vector
$\boldsymbol{k}$ being non-zero, it implies $\psi\not\equiv0$.
We summarize the results of the section in
\begin{lemma}\label{lm4.4}
The eigenvalues of $\mathcal{H}_l$ converging to a
$(p_-+p_+)$-multiple $\l_*\in\si_*$ coincide with the roots of
(\ref{4.15}) converging to $\l_*$. The associated eigenfunctions
are given by (\ref{4.3}), (\ref{4.16}), (\ref{4.20}), where the
coefficients $k_i$ form non-trivial solutions to (\ref{4.14}).
If $\l(l)$ is an eigenvalue of $\mathcal{H}_l$ converging to
$\l_*$ as $l\to+\infty$, its multiplicity coincides with the
number of linear independent solutions of the system
(\ref{4.14}) taken for $\l=\l(l)$. The associated eigenfunctions
satisfy (\ref{1.8}).
\end{lemma}

\section{Proof of Theorem~\ref{th1.3}}

Throughout this and next sections the parameter $\l$ is assumed
to belong to a small neighbourhood of $\l_*$, while $l$ is
supposed to be large enough. We begin with the proof of
Lemma~\ref{lm1.1}.
\begin{proof}[Proof of Lemma~\ref{lm1.1}]
We will prove the lemma for $\psi_i^-$ only, the case of
$\psi_i^+$ is completely similar. According to Lemma~\ref{lm3.1}
the functions $\psi_i^-$ can be represented as the series
(\ref{3.1}) in $\Pi_{a_-}^+$. Let $\Psi$ be the space of the
$L_2(\Pi)$-functions spanned over $\psi_i^-$. Each function from
this space satisfies the representation (\ref{3.1}) in
$\Pi_{a_-}^+$. We introduce two quadratic forms in this
finite-dimensional space, the first being generated by the inner
product in $L_2(\Pi)$, while the other is defined as
$\mathfrak{q}(u,v):=\a_1[u]\overline{\a_1[v]}$, where the
$\a_1[u]$, $\a_1[v]$ are the first coefficients in the
representations (\ref{3.1}) for $u$ and $v$ in $\Pi_{a_-}^+$. By
the theorem on simultaneous diagonalization of two quadratic
forms, we conclude that we can choose the basis in $\Psi$ so
that both these forms are diagonalized. Denoting this basis as
$\psi_i^-$, we conclude that these functions are orthonormalized
in $L_2(\Pi)$, and
\begin{equation}\label{6.21}
\mathfrak{q}(\psi_i^-,\psi_j^-)=0,\quad\text{if}\quad i\not=j.
\end{equation}
Suppose that for all the functions $\psi_i^-$ the coefficient
$\a_1[\psi_i^-]$ is zero. In this case we arrive at
(\ref{1.14}), where $\b_-=0$. If at least one of the functions
$\psi_i^-$ has a nonzero coefficient $\a_1$, say $\psi_1^-$, the
identity (\ref{6.21}) implies that $\a_1[\psi_i^-]=0$,
$i\geqslant2$, and we arrive again at (\ref{1.14}).
\end{proof}

The definition of $\mathrm{A}_{ij}$ and (\ref{5.2}) imply the
estimates
\begin{equation}\label{5.3}
|A_{ij}(\l,l)|\leqslant C\E^{-2ls_1(\l_*)},\quad \left|\frac{\p
A_{ij}}{\p\l}(\l,l)\right|\leqslant
Cl^2\E^{-2l(s_1(\l_*)+s_1(\l))},
\end{equation}
where the constant $C$ is independent of $\l$ and $l$. Moreover,
the holomorphy of the operator $\mathcal{T}_3$ w.r.t. $\l$
yields that the functions $A_{ij}$ are holomorphic w.r.t. $\l$.
This fact and the estimate (\ref{5.3}) allow us to claim that
the right hand side of (\ref{4.15}) reads as follows,
\begin{equation*}
F(\l,l):=\det\big((\l-\l_*)\mathrm{E}+\mathrm{A}(\l,l)\big)=
(\l-\l_*)^p+\sum\limits_{i=0}^{p-1}P_i(\l,l)(\l-\l_*)^i,
\end{equation*}
where the functions $P_i$ are holomorphic w.r.t. $\l$, and obey
the estimate
\begin{equation}\label{5.5}
|P_i(\l,l)|\leqslant C\E^{-2(p-i)l s_1(\l_*)}
\end{equation}
with the constant $C$ independent of $\l$ and $l$. Given $\d>0$,
this estimate implies
\begin{equation*}
\left|\sum\limits_{i=0}^{p-1}P_i(\l,l)(\l-\l_*)^i\right|
<|\l-\l_*|^p \quad \text{as}\quad |\l-\l_*|=\d,
\end{equation*}
if $l$ is large enough. Now we employ Rouche theorem to infer
that the function $F(\l,l)$ has the same amount of the zeroes
inside the disk $\{\l: |\l-\l_*|<\d\}$ as the function
$\l\mapsto (\l-\l_*)^p$ does. Thus, the function $F(\l,l)$ has
exactly $p$ zeroes in this disk (counting their orders), if $l$
is large enough. The number $\d$ being arbitrary, we conclude
that the equation (\ref{4.15}) has exactly $p$ roots (counting
their orders) converging to $\l_*$ as $l\to+\infty$.

\begin{lemma}\label{lm5.2}
Suppose that $\l_1(\l)$ and $\l_2(\l)$ are different roots of
(\ref{4.15}), and $\boldsymbol{k}_1(l)$ and
$\boldsymbol{k}_2(l)$ are the associated non-trivial solutions
to (\ref{4.14}) normalized by (\ref{4.19}). Then
\begin{equation*}
\big(\boldsymbol{k}_1(l),\boldsymbol{k}_2(l)\big)_{\mathbb{C}^p}=
\Odr(l\E^{-2ls_1(\l_*)}),\quad l\to+\infty.
\end{equation*}
\end{lemma}
\begin{proof}
According to Lemma~\ref{lm4.4}, the numbers
$\l_1(l)\not=\l_2(l)$ are the eigenvalues of $\mathcal{H}_l$,
and the associated eigenfunctions $\psi_i(x,l)$, $i=1,2$, are
generated by (\ref{4.3}), (\ref{4.16}), (\ref{4.20}), where
$k_i$ are components of the vectors $\boldsymbol{k}_1$,
$\boldsymbol{k}_2$, respectively.

Using the representations (\ref{3.1}) for $\psi_i^+$ in
$\Pi^\pm_{\pm a_+}$ and for $\psi_i^-$ in $\Pi^\pm_{\pm a_-}$,
by direct calculations one can check that
\begin{equation*}
\big(\psi_i^+,\mathcal{S}(2l)\psi_j^-\big)_{L_2(\Pi)}=
\Odr(l\E^{-2ls_1(\l_*)}),\quad l\to+\infty.
\end{equation*}
The operator being self-adjoint, the eigenfunctions
$\psi_i(x,l)$ are orthogonal in $L_2(\Pi)$. Now be
Lemma~\ref{lm4.4} and the last identity we obtain
\begin{equation*}
0=(\psi_1,\psi_2)_{L_2(\Pi)}=\sum\limits_{i=1}^{p} k_{i}^{(1)}
\overline{k_{i}^{(2)}}+\Odr(\E^{-2ls_1(\l_*)}),\quad
l\to+\infty,
\end{equation*}
that completes the proof.
\end{proof}

For each root $\l(l)\xrightarrow[l\to+\infty]{}\l_*$ of
(\ref{4.15}) the system (\ref{4.14}) has a finite number of
linear independent solutions. Without loss of generality we
assume that these solutions are orthonormalized in
$\mathbb{C}^p$. We consider the set of all such solutions
associated with all roots of (\ref{4.15}) converging to $\l_*$
as $l\to+\infty$, and indicate these vectors as
$\boldsymbol{k}_i=\boldsymbol{k}_i(l)$, $i=1,\ldots,q$. In view
of the assumption for $\boldsymbol{k}_i$ just made and
Lemma~\ref{lm5.2} the vectors $\boldsymbol{k}_i$ satisfy
(\ref{5.8}).

For the sake of brevity we denote
$\mathrm{B}(\l,l):=(\l-\l_*)\mathrm{E}-\mathrm{A}(\l,l)$.

\begin{lemma}\label{lm5.3}
Let $\l(l)\xrightarrow[l\to+\infty]{}\l_*$ be a root of
(\ref{4.15}) and $\boldsymbol{k}_i$, $i=N,\ldots,N+m$,
$m\geqslant0$, be the associated solutions to (\ref{4.14}). Then
for any $\boldsymbol{h}\in\mathbb{C}^p$ the representation
\begin{equation*}
\mathrm{B}^{-1}(\l,l)\boldsymbol{h}=\sum\limits_{i=N}^{N+m}
\frac{\mathcal{T}_8^{(i)}(l)\boldsymbol{h}}{\l-\l(l)}
\boldsymbol{k}_i(l)+\mathcal{T}_9(\l,l)\boldsymbol{h}
\end{equation*}
holds true for all $\l$ close to $\l(l)$. Here
$\mathcal{T}_8^{(i)}: \mathbb{C}^p\to\mathbb{C}$ are
functionals, while the matrix $\mathcal{T}_9(\l,l)$ is
holomorphic w.r.t. $\l$ in a neighbourhood of $\l(l)$.
\end{lemma}
\begin{proof}
The matrix $\mathrm{B}$ being holomorphic w.r.t. $\l$, the
inverse $\mathrm{B}^{-1}$ is meromorphic w.r.t. $\l$ and has a
pole at $\l(l)$. The residue at this pole is a linear
combination of $\boldsymbol{k}_i$, and for any
$\boldsymbol{h}\in \mathbb{C}^p$ we have
\begin{equation}\label{5.10}
\mathrm{B}^{-1}(\l,l)\boldsymbol{h} = \frac{1}{(\l-\l(l))^s}
\sum\limits_{i=N}^{N+m} \boldsymbol{k}_i
\mathcal{T}_8^{(i)}(l)\boldsymbol{h} +\Odr
\big((\l-\l(l))^{-s+1}\big),\quad \l\to\l(l),
\end{equation}
where $s\geqslant 1$ is the order of the pole, and
$\mathcal{T}_8^{(i)}:\mathbb{C}^p\to \mathbb{C}$ are some
functionals.

We are going to prove that $s=1$. Let $g_\pm=g_\pm(x)\in
L_2(\Pi,\Om_\pm)$ be arbitrary functions. We consider the
equation
\begin{equation}\label{5.12}
(\mathcal{H}_l-\l)u=g:=\mathcal{S}(-l)g_+ +\mathcal{S}(l)g_-,
\end{equation}
where $\l$ is close to $\l(l)$ and $\l\not=\l(l)$,
$\l\not=\l_*$. The results of \cite[Ch. V\!I\!I, Sec. 3.5]{K}
imply that for such $\l$ the function $u$ can be represented as
\begin{equation}\label{5.12a}
u=-\frac{1}{\l-\l(l)}\sum\limits_{i=N}^{N+m}(g,\psi_i)_{L_2(\Pi)}
\psi_i+\mathcal{T}_{10}(\l,l)g,
\end{equation}
where $\psi_i$ are the eigenfunctions of $\mathcal{H}_l$
associated with $\l(l)$ and orthonormalized in $L_2(\Pi)$, while
the operator $\mathcal{T}_{10}(\l,l)$ is bounded and holomorphic
w.r.t. $\l$ close to $\l(l)$ as an operator in $L_2(\Pi)$. The
eigenvalue $\l(l)$ of the operator $\mathcal{H}_l$ is
$(m+1)$-multiple by the assumption and Lemma~\ref{lm4.4}.
Completely by analogy with the proof of Lemma~\ref{lm4.2} one
can make sure that the equation (\ref{5.12}) is equivalent to
\begin{equation}\label{5.13}
\boldsymbol{f}+\mathcal{T}_2(\l,l)\boldsymbol{f}=\boldsymbol{g},
\end{equation}
where $\boldsymbol{g}:=(g_-,g_+)\in L_2(\Om_-)\oplus
L_2(\Om_+)$, and the solution of (\ref{5.12}) is given by
\begin{equation}\label{5.14}
u=\mathcal{S}(l)u_-+\mathcal{S}(-l)u_+,\quad
u_\pm=(\mathcal{H}_\pm-\l)^{-1}f_\pm.
\end{equation}
Proceeding as in (\ref{4.5}), (\ref{4.8a}), (\ref{4.13}), one
can solve (\ref{5.13}),
\begin{equation}\label{5.17}
\boldsymbol{f}=\sum\limits_{i=1}^{p}U_i\boldsymbol{\Phi}_i+
\widetilde{\boldsymbol{f}},\quad
\boldsymbol{U}=\mathrm{B}^{-1}(\l,l)\boldsymbol{h},\quad
\widetilde{\boldsymbol{f}}:=(\I+\mathcal{T}_3)^{-1}
\boldsymbol{g},
\end{equation}
where $U_i:=\mathcal{T}_1^{(i)}\boldsymbol{f}$, and the vector
$\boldsymbol{U}:=(U_1 \ldots U_p)^{t}$ is a solution to
\begin{gather}
\mathrm{B}(\l,l)\boldsymbol{U}=\boldsymbol{h},\label{5.12b}
\\
\boldsymbol{h}= (h_1 \ldots h_p)^{t},\quad h_i:=
\mathcal{T}_1^{(i)}(\l)(\I+\mathcal{T}_3(\l,l))^{-1}\boldsymbol{g}
\label{5.16}
\end{gather}
Knowing the vectors $\boldsymbol{U}$ and $\boldsymbol{f}$, we
can restore the functions $u_\pm$  by  (\ref{3.4}),
\begin{align*}
&u_-(\cdot,\l,l)=-\sum\limits_{i=1}^{p_-}U_i(\l,l)\psi_i^--
\sum\limits_{i=1}^{p}U_i(\l,l) \mathcal{T}_4^-(\l,l)\Phi_i^-+
\mathcal{T}_4^-(\l)\widetilde{f}_-,
\\
&u_+(\cdot,\l,l)=-\sum\limits_{i=1}^{p_+}U_{i+p_-}(\l,l)\psi_i^+-
\sum\limits_{i=1}^{p}U_i(\l,l) \mathcal{T}_4^+(\l,l)\Phi_i^++
\mathcal{T}_4^+(\l)\widetilde{f}_+,
\end{align*}
where $\Phi_i^\pm$ and $\widetilde{f}_\pm$ are introduced as
$\boldsymbol{\Phi}_i=(\Phi_i^-,\Phi_i^+)$,
$\widetilde{\boldsymbol{f}}=(\widetilde{f}_-,\widetilde{f}_+)$.
In these formulas we have also employed the system (\ref{5.12b})
in the following way:
\begin{align*}
&(f_-,\psi_j^-)_{L_2(\Pi)}=\mathcal{T}_1^{(j)}\boldsymbol{f} =
h_j+\sum\limits_{i=1}^{p}A_{ji}U_i=(\l-\l_*)U_j,\quad
j=1,\ldots,p_-,
\\
&(f_+,\psi_j^+)_{L_2(\Pi)}=\mathcal{T}_1^{(j+p_-)}\boldsymbol{f}
= h_{j+p_-}+\sum\limits_{i=1}^{p} A_{ji}U_i
=(\l-\l_*)U_{j+p_-},\quad j=1,\ldots,p_+.
\end{align*}
The estimates (\ref{5.2}) allow us to infer that
\begin{equation*}
\|\mathcal{T}_4^\pm(\l,l)\Phi_i^\pm\|_{L_2(\Pi)}=
\Odr(\E^{-2ls_1(\l_*)}),
\end{equation*}
while in view of holomorphy of $\mathcal{T}_4^\pm$ and
(\ref{4.12}) we have
\begin{equation*}
\|\mathcal{T}_4^\pm(\l)\widetilde{f}_\pm\|_{L_2(\Pi)}\leqslant
C\|\boldsymbol{g}\|_{L_2(\Om_-)\oplus L_2(\Om_+)},
\end{equation*}
where the constant $C$ is independent of $\l$. Now we use the
first of the relations (\ref{5.14}) and obtain
\begin{equation}\label{5.18}
\begin{aligned}
u&(\cdot,\l,l)=-\sum\limits_{i=1}^{p_-}U_i(\l,l) \Big(
\mathcal{S}(l)\psi_i^- + \Odr(\E^{-2ls_1(\l_*)})\Big)
\\
&-\sum\limits_{i=1}^{p_+}U_{i+p_-}(\l,l)
\Big(\mathcal{S}(-l)\psi_i^+ + \Odr(\E^{-2ls_1(\l_*)})\Big)+
\Odr\left(\|\boldsymbol{g}\|_{L_2(\Om_-)\oplus
L_2(\Om_+)}\right).
\end{aligned}
\end{equation}

Now we compare the representations (\ref{5.12a}) and
(\ref{5.18}) and conclude that the coefficients $U_i(\l,l)$,
$i=1,\ldots,p$ has a simple pole at $\l(l)$. By (\ref{5.17}) it
implies that the vector $\mathrm{B}^{-1}(\l,l)\boldsymbol{h}$
has a simple pole at $\l(l)$, where $\boldsymbol{h}$ is defined
by (\ref{5.16}). It follows from (\ref{4.12}) and (\ref{5.16})
that for each $\boldsymbol{h}\in\mathbb{C}^p$ there exists
$\boldsymbol{g}\in L_2(\Om_-)\oplus L_2(\Om_+)$ so that
$\boldsymbol{h}$ is given by (\ref{5.16}). Together with
(\ref{5.10}) it completes the proof.
\end{proof}

\begin{lemma}\label{lm5.1}
The number $\l(l)\xrightarrow[l\to+\infty]{}\l_*$ is a $m$-th
order zero of $F(\l,l)$ if and only if it is a $m$-multiple
eigenvalue of  $\mathcal{H}_l$.
\end{lemma}
\begin{proof}
Let $\l^{(i)}(l)\xrightarrow[l\to+\infty]{}\l_*$,
$i=1,\ldots,M$, be the different zeroes of $F(\l,l)$, and $r_i$,
$i=1,\ldots,M$, be the orders of these zeroes. By
Lemma~\ref{lm4.4}, each zero $\l^{(i)}(l)$ is an eigenvalue of
$\mathcal{H}_l$; its multiplicity will be indicated as
$m_i\geqslant 1$. To prove the lemma it is sufficient to show
that $m_i=r_i$, $i=1,\ldots,M$.

Let us prove first that $m_i\leqslant r_i$. In accordance with
Lemma~\ref{lm4.4} the multiplicity $m_i$ coincides with a number
of linear independent solutions of (\ref{4.14}) with
$\l=\l^{(i)}(l)$. Hence,
\begin{equation}\label{5.21}
\rank \mathrm{B}(\l(l),l)=p-m_i.
\end{equation}
By the assumption
\begin{equation*}
\frac{\p^j}{\p\l^j}\det \mathrm{B}(\l,l)=0,\quad
j=1,\ldots,r_i-1,\qquad \frac{\p^{r_i}}{\p\l^{r_i}}\det
\mathrm{B}(\l,l)\not=0,
\end{equation*}
as $\l=\l^{(i)}(l)$. Let $B_j=B_j(\l,l)$ be the columns of the
matrix $\mathrm{B}$, i.e., $\mathrm{B}=(B_1,\ldots,B_p)$.
Employing the well-known formula
\begin{equation*}
\frac{\p}{\p\l}\det \mathrm{B}=\det\left(\frac{\p
B_1}{\p\l}\,B_2\,\ldots\,B_p\right) + \det\left(B_1\,\frac{\p
B_2}{\p\l}\,\ldots\,B_p\right)+
\det\left(B_1\,B_2\,\ldots\,\frac{\p  B_p}{\p\l}\right),
\end{equation*}
one can check easily that for each $0\leqslant j\leqslant m_i-1$
\begin{equation*}
\frac{\p^j}{\p\l^j}\det \mathrm{B}(\l,l)\Big|_{\l=\l^{(i)}(l)}
=\sum\limits_{\vs} c_\vs\det \mathrm{B}_\vs,
\end{equation*}
where $c_\vs$ are constants, and at least $(p-m_i+1)$ columns of
each matrix $\mathrm{B}_\vs$ are those of $\mathrm{B}$. In view
of (\ref{5.21}) these columns are linear dependent, and
therefore $\det \mathrm{B}_\vs=0$ for each $\vs$. Thus
$m_i-1\geqslant r_i-1$ that implies the desired inequality.

Now it is sufficient to check that
$q=\sum\limits_{i=1}^{M}m_i=\sum\limits_{i=1}^{M}r_i=p$ to prove
that $m_i=r_i$. Lemma~\ref{lm5.3} yields that for a given fixed
$\d$ small enough and $l$ large enough
\begin{equation}\label{5.20a}
\mathrm{B}^{-1}(\l,l)\boldsymbol{h}=\sum\limits_{i=1}^{q}
\frac{\mathcal{T}_8^{(i)}(l)\boldsymbol{h}}{\l-\l^{(i)}(l)}
\boldsymbol{k}_i(l)+\mathcal{T}_9(\l,l)\boldsymbol{h},
\end{equation}
for any $\boldsymbol{h}\in\mathbb{C}^p$ and $\l$ so that
$|\l-\l_*|=\d$. It is also assumed that $|\l^{(i)}(l)-\l_*|<\d$
for the considered values of $l$. We integrate this identity to
obtain
\begin{equation}\label{5.22}
\frac{1}{2\pi\iu}\int\limits_{|\l-\l_*|=\d}
\mathrm{B}^{-1}(\l,l_s)\boldsymbol{h}\di\l
=\sum\limits_{i=1}^{q}\boldsymbol{k}_i
\mathcal{T}_8^{(i)}(l)\boldsymbol{h}.
\end{equation}
Due to (\ref{5.3}) we conclude that
\begin{equation}\label{5.23}
\frac{1}{2\pi\iu}\int\limits_{|\l-\l_*|=\d}
\mathrm{B}^{-1}(\l,l)\di\l\xrightarrow[l\to+\infty]{}
\frac{1}{2\pi\iu}\int\limits_{|\l-\l_*|=\d}
\frac{\mathrm{E}\di\l}{\l-\l_*}=\mathrm{E}.
\end{equation}
Hence the right hand side of (\ref{5.22}) converges to
$\boldsymbol{h}$. By (\ref{5.8}) it implies that $q=p$ for $l$
large enough.
\end{proof}

The statement of Theorem~\ref{th1.3} follows from the proven
lemma.

\section{Asymptotics for the eigenelements of  $\mathcal{H}_l$}

In this section we prove Theorems~\ref{th1.4}-\ref{th1.8}.
Throughout the section  the hypothesis of Theorem~\ref{th1.4} is
assumed to hold true.

Theorem~\ref{th1.3} implies that the number of the vectors
$\boldsymbol{k}_i$ introduced in the previous section equals
$p$. Let $\mathrm{S}=\mathrm{S}(l)$ be the matrix with columns
$\boldsymbol{k}_i(l)$, $i=1,\ldots,p$, i.e.,
$\mathrm{S}(l)=(\boldsymbol{k}_1(l)\ldots\boldsymbol{k}_p(l))$.
Without loss of generality we can assume that $\det
\mathrm{S}(l)\geqslant0$.

\begin{lemma}\label{lm6.1}
$\det\mathrm{S}(l)=1+\Odr(l\E^{-2ls_1(\l_*)})$, as
$l\to+\infty$.
\end{lemma}
\begin{proof}
The relations (\ref{5.8}) yield
$\mathrm{S}^2(l)=\mathrm{E}+\Odr(l\E^{-2ls_1(\l_*)})$,
$l\to+\infty$, that implies
$\det^2\mathrm{S}(l)=1+\Odr(l\E^{-2ls_1(\l_*)})$. The last
identity proves the lemma.
\end{proof}

Lemma~\ref{lm6.1} implies that there exists the inverse matrix
$\mathrm{S}^{-1}(l)$ for $l$ large enough.

\begin{lemma}\label{lm6.2}
The matrix
$\mathrm{R}(\l,l):=\mathrm{S}^{-1}(l)\mathrm{A}(\l,l)\mathrm{S}(l)$
reads as follows
\begin{equation*}
\mathrm{R}=
\begin{pmatrix}
\l_1-\l_*+(\l-\l_1)r_{11} & (\l-\l_2)r_{12} & \ldots &
(\l-\l_p)r_{1p}
\\
(\l-\l_1)r_{21} & \l_2-\l_*+(\l-\l_2)r_{22}& \ldots &
(\l-\l_p)r_{2p}
\\
\vdots & \vdots & \hphantom{\ldots} & \vdots
\\
(\l-\l_1)r_{p1} & (\l-\l_2)r_{p2} & \ldots &
\l_p-\l_*+(\l-\l_p)r_{pp}
\end{pmatrix}.
\end{equation*}
where $\l_i=\l_i(l)$, while the functions $r_{ij}=r_{ij}(\l,l)$
are holomorphic w.r.t. $\l$ close to $\l_*$ and obey the uniform
in $\l$ and $l$ estimates
\begin{equation}\label{6.2}
|r_{ij}(\l,l)|\leqslant Cl^2\E^{-2l(s_1(\l_*)+s_1(\l))}.
\end{equation}
\end{lemma}

\begin{proof}
The system (\ref{4.14}) implies
\begin{align*}
\mathrm{A}(\l,l)\boldsymbol{k}_i&=
\mathrm{A}(\l_i(l),l)\boldsymbol{k}_i+
\big(\mathrm{A}(\l,l)-\mathrm{A}(\l_i(l),l)\big)
\boldsymbol{k}_i
\\
&=(\l_i(l)-\l_*)\boldsymbol{k}_i+\big(\mathrm{A}(\l,l)-\mathrm{A}(\l_i(l),l)\big)
\boldsymbol{k}_i.
\end{align*}
The matrix $\mathrm{A}(\l,l)-\mathrm{A}(\l_i(l),l)$ is
holomorphic w.r.t. $\l$, and
\begin{equation}\label{6.1a}
\widetilde{\mathrm{A}}(\l,\l_i(l),l):=
\mathrm{A}(\l,l)-\mathrm{A}(\l_i(l),l)=\int\limits_{\l_i(l)}^{\l}
\frac{\p \mathrm{A}}{\p\l}(z,l)\di z.
\end{equation}
Due to (\ref{5.3}) and (\ref{4.14}) the last identity implies
that
\begin{equation}\label{6.5}
\mathrm{A}(\l,l)\boldsymbol{k}_i=(\l_i(l)-\l_*)\boldsymbol{k}_i
+ (\l-\l_i)\boldsymbol{K}_i(\l,l),
\end{equation}
where the vectors $\boldsymbol{K}_i(\l,l)$ are holomorphic
w.r.t. $\l$ close to $\l_*$ and satisfy the uniform in $\l$ and
$l$ estimate
\begin{equation*}
\|\boldsymbol{K}_i\|_{\mathbb{C}^p}\leqslant
Cl^2\E^{-2l(s_1(\l_*)+s_1(\l))}.
\end{equation*}
By Lemma~\ref{lm5.1} the vectors $\boldsymbol{k}_i$,
$i=1,\ldots,p$, form a basis in $\mathbb{C}^p$. Hence,
\begin{equation}
\boldsymbol{K}_i(\l,l)=
\sum\limits_{j=1}^{p}r_{ij}(\l,l)\boldsymbol{k}_j(l),\quad \big(
r_{i1}(\l,l) \ldots r_{ip}(\l,l)\big)^{t}=
\mathrm{S}^{-1}(l)\boldsymbol{K}_i(\l,l).\label{6.5a}
\end{equation}
Due to Lemma~\ref{lm6.1}, the relations (\ref{5.8}), and the
established properties of $\boldsymbol{K}_i$ we infer that the
functions $r_{ij}$ are holomorphic w.r.t. $\l$ and satisfy \
(\ref{6.2}). Taking into account  (\ref{6.5}) and (\ref{6.5a}),
we arrive at the statement of the lemma.
\end{proof}

\begin{lemma}\label{lm6.3}
The polynomial $\det\big(\tau\mathrm{E}-\mathrm{A}(\l_*,l)\big)$
has exactly $p$ roots $\tau_i=\tau_i(l)$, $i=1,\ldots,p$
counting multiplicity which satisfy (\ref{1.6}).
\end{lemma}
\begin{proof}
It is clear that
$\det\big(\tau\mathrm{E}-\mathrm{A}(\l_*,l)\big)$ is a
polynomial of $p$-th order, this is why it has $p$ roots
$\tau_i(l)$, $i=1,\ldots,p$, counting multiplicity. It is easy
to check that
\begin{equation*}
\det\big(\tau\mathrm{E}-\mathrm{A}(\l_*,l)\big)=\tau^p
+\sum\limits_{j=0}^{p-1}P_j(\l_*,l)\tau^j,
\end{equation*}
where the functions $P_j(\l_*,l)$ satisfy (\ref{5.5}). We make a
change of variable $\tau_i=z_i\E^{-2ls_1(\l_*)}$, and together
with the representation just obtained it leads us to the
equation for $z_i$,
\begin{equation}\label{6.5b}
z^p+\sum\limits_{j=0}^{p-1}\E^{-2l(j-p)s_1(\l_*)}P_j(\l_*,l)
z^j=0.
\end{equation}
Due to (\ref{5.5}) the coefficients of this equation are bounded
uniformly in $l$. By Rouche theorem it implies that all the
roots of (\ref{6.5b}) are bounded uniformly in $l$. This fact
leads us to (\ref{1.6}).
\end{proof}

In what follows the roots $\tau_i$ are supposed to be ordered in
accordance with (\ref{1.7}). We denote $\mu_i(l):=\l_i(l)-\l_*$,
$i=1,\ldots,p$.

\begin{proof}[Proof of Theorem~\ref{th1.4}]
The formulas (\ref{1.6}), (\ref{1.8}) were established in
Lemma~\ref{lm6.3}. Let us prove (\ref{1.5}). Namely, let us
prove that for each $l$ large enough the roots of (\ref{4.15})
can be ordered so that
\begin{equation}\label{6.6}
\tau_i(l)=\mu_i(l)\left(1+\Odr\left(l^{\frac{2}{p}}
\E^{-\frac{4l}{p}s_1(\l_*)}\right)\right),\quad l\to+\infty.
\end{equation}
Assume that this not true on a sequence $\l_{s}\to+\infty$. We
introduce an equivalence relation $\sim$ on
$\{\mu_{i}(l_{s})\}_{i=1,\ldots,p}$ saying that
$\mu_i\sim\mu_j$, if
\begin{equation*}
\mu_i(l_{s})=\mu_j(l_{s})\left(1+\Odr\left(l_{s}^{\frac{2}{p}}
\E^{-\frac{4l_{s}}{p}s_1(\l_*)}\right)\right),\quad
l_{s}\to+\infty.
\end{equation*}
This relation divides all $\mu_i(l_s)$ into disjoint groups,
\begin{equation*}
\{\l_1(l),\ldots,\l_p(l)\}=\bigcup\limits_{i=1}^q
\{\l_{m_i}(l),\ldots, \l_{m_{i+1}-1}(l)\},
\end{equation*}
where $1=m_1<m_2<\ldots<m_{q+1}=p+1$, $\l_k \sim\l_t$,
$k,t=m_i,\ldots,m_{i+1}-1$, $i=1,\ldots,q$, and
$\l_k\not\sim\l_t$, if $m_i\leqslant k\leqslant m_{i+1}-1$,
$m_j\leqslant t\leqslant m_{j+1}-1$, $i\not=j$. Extracting if
needed a subsequence from $\{l_s\}$, we assume that $m_i$ and
$q$ are independent of $\{l_s\}$. Given $k\in\{1,\ldots,p\}$,
there exists $i$ such that $m_i\leqslant k\leqslant m_{i+1}-1$.
For the sake of brevity we denote $m:=m_i$,
$\widetilde{m}:=m_{i+1}$, $\widehat{m}:=\widetilde{m}-m$. To
prove (\ref{1.5}) it is sufficient to show that $\widehat{m}$
roots $\tau_i$, $i=m,\ldots,\widetilde{m}-1$, counting
multiplicity of
$\det\big(\tau\mathrm{E}-\mathrm{A}(\l_*,l)\big)$ satisfy
(\ref{6.6}).

Since
\begin{equation*}
\det\big(\tau\mathrm{E}-\mathrm{A}(\l_*,l)\big)=
\det\big(\mathrm{S}^{-1}(l)(\tau\mathrm{E}-\mathrm{A}(\l_*,l))
\mathrm{S}(l)\big)=\det\big(\tau\mathrm{E}-\mathrm{R}(\l_*,l)\big),
\end{equation*}
due to Lemma~\ref{lm6.2} the equation for $\tau_i$ can be
rewritten as
\begin{equation}\label{6.7}
\begin{vmatrix}
\tau-\mu_1(1-r_{11}) & -\mu_2r_{12} & \ldots & -\mu_p r_{1p}
\\
-\mu_1r_{21} & \tau-\mu_2(1-r_{22}) &\ldots & -\mu_p r_{2p}
\\
\vdots & \vdots &  & \vdots
\\
-\mu_1r_{p1} & -\mu_2r_{p2} & \ldots & \tau-\mu_p(1-r_{pp})
\end{vmatrix}
=0,
\end{equation}
where $\mu_j=\mu_j(l)$, $r_{jk}=r_{jk}(\l_*,l)$. Assume first
that $\mu_k(l_s)=0$. In view of (\ref{1.4}) it implies that
$m=m_1$, $\widetilde{m}=m_2$, and $\mu_j(l_s)=0$,
$j=m,\ldots,\widetilde{m}-1$. In this case the equation
(\ref{6.7}) becomes
\begin{equation*}
\begin{vmatrix}
\tau &  \ldots & 0 & -\mu_{\widetilde{m}} r_{1\widetilde{m}} &
\ldots & -\mu_p r_{1p}
\\
\vdots  &  & \vdots & \vdots && \vdots
\\
0  & \ldots &\tau & -\mu_{\widetilde{m}}r_{\widetilde{m}-1
\widetilde{m}} & \ldots & -\mu_p r_{\widetilde{m}-1p}
\\
\vdots  &  & \vdots  & \vdots && \vdots
\\
0  & \ldots & 0 &  -\mu_{\widetilde{m}}r_{p\widetilde{m}} &
\ldots & \tau-\mu_p(1-r_{pp})
\end{vmatrix}=0,
\end{equation*}
and it implies that zero is the root of $\det\big(\tau\mathrm{E}
-\mathrm{A}(\l_*,l)\big)$ of multiplicity at least
$\widehat{m}$. In this case the identities (\ref{6.6}) are
obviously valid.

Assume now that $\mu_k(l_s)\not=0$. We seek the needed roots as
$\tau=\mu_k(l_s)(1+z)$. We substitute this identity into
(\ref{6.7}) and divide then first $(\widetilde{m}-1)$ columns by
$-\mu_k(l_s)$, while the other columns are divided by the
functions $-\mu_j(l_s)$ corresponding to them. This procedure
leads us to the equation
\begin{gather*}
\begin{vmatrix}
z_1-z & \ldots &
\frac{\mu_{\widetilde{m}-1}}{\mu_k}r_{1\widetilde{m}-1} &
r_{1\widetilde{m}} &  \ldots & r_{1p}
\\
\vdots & & \vdots & \vdots & & \vdots
\\
\frac{\mu_1}{\mu_k}r_{\widetilde{m}-11} & \ldots &
z_{\widetilde{m}-1}-z & r_{\widetilde{m}-1\widetilde{m}} &
\ldots & r_{\widetilde{m}-1p}
\\
\frac{\mu_1}{\mu_k}r_{\widetilde{m}1} & \ldots &
\frac{\mu_{\widetilde{m}-1}}{\mu_k}r_{\widetilde{m}\widetilde{m}-1}
& z_{\widetilde{m}}-\frac{\mu_k}{\mu_{\widetilde{m}}}z & \ldots
& r_{\widetilde{m}p}
\\
\vdots & & \vdots & \vdots & & \vdots
\\
\frac{\mu_1}{\mu_k}r_{p1} & \ldots &
\frac{\mu_{\widetilde{m}-1}}{\mu_k}r_{p\widetilde{m}-1} &
r_{p\widetilde{m}} & \ldots & z_p-\frac{\mu_k}{\mu_p}z
\end{vmatrix}
=0,
\\
\
\\
\begin{aligned}
&z_i=z_i(l):=\frac{\mu_i(l)}{\mu_k(l)}(1-r_{ii}(\l_*,l))-1,&&
i=1,\ldots,\widetilde{m}-1,
\\
&z_i=z_i(l):=1-\frac{\mu_k(l)}{\mu_i(l)}-r_{ii}(\l_*,l),&&
i=\widetilde{m},\ldots,p,
\end{aligned}
\end{gather*}
and the arguments of all the functions are $l=l_s$, $\l=\l_*$.
Due to (\ref{1.4}) all the fractions in this determinant are
bounded uniformly in $l_s$. Using this fact and (\ref{6.2}), we
calculate this determinant and write the multiplication of the
diagonal separately,
\begin{align}\label{6.8a}
&F_1(z,l_s)F_2(z,l_s)F_3(z,l_s)-F_4(z,l_s)=0,
\\
& F_1(z,l_s):=\prod\limits_{i=1}^{m-1}\left(z_i(l_s)-z\right),
\quad F_2(z,l_s):=\prod\limits_{i=m}^{\widetilde{m}-1}
\left(z_i(l_s)-z\right),\nonumber
\\
&F_3(z,l_s):=\prod\limits_{i=\widetilde{m}}^p \left(z_i(l_s)-
\frac{\mu_i(l_s)}{\mu_k(l_s)}z\right),\quad
F_4(z,l_s):=\sum\limits_{i=0}^{p-2}z^i c_i(l_s)\nonumber
\end{align}
where the coefficients $c_i$ obey the estimate
\begin{equation}\label{6.9}
c_i(l_s)=\Odr\Big(l_s^2\E^{-4l_s s_1(\l_*)}\Big),\quad
l_s\to+\infty.
\end{equation}
As it follows from the definition of the equivalence relation,
\begin{align*}
&\frac{|\mu_i-\mu_k|}{|\mu_k|}\leqslant C_0\z,\quad
l_{s}\to+\infty, \quad i=m,\ldots,\widetilde{m}-1,\quad
\z=\z(l_s):=l_s^2 \E^{-4l_s s_1(\l_*)},
\\
&\frac{|\mu_i-\mu_k|}{|\mu_k|}\geqslant \th\z,\quad
i\not\in\{m,\ldots,\widetilde{m}-1\},\quad
\th=\th(l_s)\to+\infty,\quad l_{s}\to+\infty,
\end{align*}
where the constant $C_0$ is independent of $l_s$. These
estimates and (\ref{6.2}), (\ref{1.4}) imply
\begin{equation}\label{6.10b}
|F_1(z,l)|\geqslant C(\th\z)^{m-1},\quad |F_3(z,l)|\geqslant
C(\th\z)^{p-\widetilde{m}+1},\quad |z|\leqslant 2C_0\z.
\end{equation}
The zeroes $z_i(l_s)$ of $F_2(z,l_s)$ satisfy
$|z_i(l_s)|\leqslant C_0\z$, and hence $|F_2(z,l_s)|\geqslant
C_0^{\widehat{m}}\z^{\widehat{m}}$ as $|z|=C_0\z$. Employing
this estimate, (\ref{6.10b}) and rewriting (\ref{6.8a}) as
\begin{equation*}
F_2(z,l_s)-\frac{F_4(z,l_s)}{F_1(z,l_s)F_3(z,l_s)}=0,
\end{equation*}
by Rouche theorem we conclude that the last equation has exactly
$\widehat{m}$ roots counting multiplicity in the disk
$\{|z|<2C_0\z\}$. We denote these roots as $z^{(j)}(l_s)$,
$j=m,\ldots,\widetilde{m}-1$. It follows from (\ref{6.8a}),
(\ref{6.9}), (\ref{6.10b}) that
\begin{align*}
\left(\min\limits_{i=m,\ldots,\widetilde{m}-1}
|z^{(j)}-z_i|\right)^{\widehat{m}}&\leqslant
|F_2(z^{(j)},l_s)|=\left|\frac{F_4(z^{(j)},l_s)}
{F_1(z^{(j)},l_s)F_3(z^{(j)},l_s)}\right|
\\
&\leqslant
C\frac{\z^{\widehat{m}}(l_s)}{\th^{p-\widehat{m}}(l_s)}\leqslant
C\z^{\widehat{m}}(l_s),
\end{align*}
where $z^{(j)}=z^{(j)}(l_s)$, and the constant $C$ is
independent of $l_s$. Hence, for each $j$ there exists index
$i$, depending on $l_s$, such that
\begin{equation*}
z^{(j)}=z_{i}+\Odr(\z),\quad l_s\to+\infty.
\end{equation*}
This identity, the definition of the equivalence relation and
(\ref{6.2}) imply
\begin{align*}
\tau^{(j)}=\mu_k(1+z^{(j)})=\mu_j(1-r_{jj})+\Odr(\mu_k\z)=
\mu_j+\Odr(\mu_j \z),
\end{align*}
which yields (\ref{6.6}).
\end{proof}

In the proof of Theorem~\ref{th1.5} we will employ the following
lemma.

\begin{lemma}\label{lm6.5}
For $\l$ close to $\l_*$ and $\boldsymbol{h}\in \mathbb{C}^p$
the representation
\begin{equation*}
\mathrm{B}^{-1}(\l,l)\boldsymbol{h}=\sum\limits_{i=1}^{p}
\frac{\mathcal{T}_{10}^{(i)}(\l,l)\boldsymbol{h}}{\l-\l_i(l)}
\boldsymbol{k}_i
\end{equation*}
holds true. Here $\mathcal{T}_{10}^{(i)}(\l,l): \mathbb{C}^p\to
\mathbb{C}$ are functionals bounded uniformly in $\l$ and $l$.
\end{lemma}

\begin{proof}
Lemma~\ref{lm5.3} implies that for $\l$ close to $\l_*$ the
identity (\ref{5.20a}) holds true for any
$\boldsymbol{h}\in\mathbb{C}^p$, where $l_s$ should be replaced
by $l$, $q=p$ and $\l^{(i)}=\l_i$. We introduce the vectors
\begin{equation*}
\boldsymbol{k}_i^\perp(l):=\frac{\widetilde{\boldsymbol{k}}_i(l)}
{\|\widetilde{\boldsymbol{k}}_i(l)\|}, \quad
\widetilde{\boldsymbol{k}}_i:=\boldsymbol{k}_i-
\sum\limits_{\genfrac{}{}{0pt}{2}
{j=1}{j\not=i}}^{p}(\boldsymbol{k}_i,\boldsymbol{k}_j)_{\mathbb{C}^p}
\boldsymbol{k}_j.
\end{equation*}
The relations (\ref{5.8}) implies that the vectors
$\boldsymbol{k}_i^\perp$ satisfy these relation as well.
Moreover, the vectors $\boldsymbol{k}_i^\perp$ form the
orthogonal basis for the basis $\{\boldsymbol{k}_i\}$. Bearing
this fact in mind, we multiply the relation (\ref{5.22}) by
$\boldsymbol{k}_j^\perp$ with $l_s$ replaced by $l$, $q=p$ and
obtain
\begin{equation*}
\frac{1}{2\pi\iu}\left(\int\limits_{|\l-\l_*|=\d}
\mathrm{B}^{-1}(\l,l)\boldsymbol{h}\di\l,
\boldsymbol{k}_j^\perp\right)_{\mathbb{C}^p} =
\mathcal{T}_8^{(j)}(l)\boldsymbol{h}.
\end{equation*}
Due to (\ref{5.23}) we conclude that the functionals
$\mathcal{T}_8^{(j)}$ are bounded uniformly in $l$.

Let us prove that the matrix $\mathcal{T}_9(\l,l)$ is bounded
uniformly in $\l$ and $l$. Due to (\ref{5.23}) and the
convergences $\l_i(\l)\to\l_*$ we have
\begin{equation*}
\|\mathcal{T}_9(\l,l)\boldsymbol{h}\|_{\mathbb{C}^p}=
\left\|\mathrm{B}^{-1}(\l,l)\boldsymbol{h}-\sum\limits_{i=1}^{p}
\frac{\mathcal{T}_8^{(i)}(l)\boldsymbol{h}}{\l-\l_i(l)}
\boldsymbol{k}_i(l)\right\|_{\mathbb{C}^p}\leqslant
C\|\boldsymbol{h}\|_{\mathbb{C}^p},
\end{equation*}
as $|\l-\l_*|=\d$, if $l$ is large enough. The constant $C$ here
is independent of $\boldsymbol{h}$ and $\l$ such that
$|\l-\l_*|=\d$. The matrix $\mathcal{T}_9$ being holomorphic
w.r.t. $\l$, by the maximum principle for holomorphic functions
the estimate holds true for $|\l-\l_*|<\d$, too. Thus, the
matrix $\mathcal{T}_9$ is bounded uniformly in $\l$ and $l$. We
can expand $\mathcal{T}_9\boldsymbol{h}$ in terms of the basis
$\{\boldsymbol{k}_i\}$,
\begin{equation}\label{6.15}
\mathcal{T}_9(\l,l)\boldsymbol{h}=\sum\limits_{i=1}^{p}
\boldsymbol{k}_i \mathcal{T}_{11}^{(i)}(\l,l)\boldsymbol{h},
\quad \left(\mathcal{T}_{11}^{(1)}\boldsymbol{h} \ldots
\mathcal{T}_{11}^{(p)}\boldsymbol{h}\right)^{t}
=\mathrm{S}^{-1}(l)\mathcal{T}_9(\l,l)\boldsymbol{h},
\end{equation}
where the functionals
$\mathcal{T}_{11}^{(i)}:\mathbb{C}^p\to\mathbb{C}$ are bounded
uniformly in $\l$ and $l$. Substituting (\ref{6.15}) into
(\ref{5.20a}) with $l_s$ replaced by $l$, $q=p$,
$\l^{(i)}=\l_i$, we arrive at the desired representation, where
$\mathcal{T}_{10}^{(i)}(\l,l)=
\mathcal{T}_8^{(i)}(l)+(\l-\l_i(l))\mathcal{T}_{11}^{(i)}(\l,l)$.
\end{proof}

\begin{proof}[Proof of Theorem~\ref{th1.5}]
Since the matrix $\mathrm{A}_0(l)$ satisfies the condition (A),
it has $p$ eigenvalues $\tau_i^{(0)}$, $i=1,\ldots,p$, counting
multiplicity. By $\boldsymbol{h}_i=\boldsymbol{h}_i(l)$,
$i=1,\ldots,p$, we denote the associated eigenvectors normalized
in $\mathbb{C}^p$. Completely by analogy with the proof of
(\ref{1.6}) in Lemma~\ref{lm6.3} one can establish (\ref{1.13}).
It is easy to check that the vectors $\boldsymbol{h}_i$ satisfy
the identities
\begin{equation*}
\mathrm{B}(\l_*+\tau_i^{(0)},l)\boldsymbol{h}_i(l)=
-\mathrm{A}_1(l)\boldsymbol{h}_i(l)-\tau_i^{(0)}
\widetilde{\mathrm{A}}(\l_*,\l_*+\tau_i^{(0)},l)\boldsymbol{h}_i(l):=
\widetilde{\boldsymbol{h}}_i(l),
\end{equation*}
where, we remind, the matrix $\widetilde{\mathrm{A}}$ was
introduced in (\ref{6.1a}). Now we employ (\ref{5.3}) to obtain
\begin{equation}\label{6.16}
\|\widetilde{\boldsymbol{h}}_i(l)\|_{\mathbb{C}^p}\leqslant
\|\mathrm{A}_1(l)\|+C|\tau_i^{(0)}|l^2\E^{-2l(s_1(\l_*)+
s_1(\widetilde{\l}_i(l)))} \leqslant
\|\mathrm{A}_1(l)\|+C|\tau_i^{(0)}|l^2\E^{-4ls_1(\l_*)},
\end{equation}
where the constant $C$ is independent of $\l$ and $l$. Here we
have also used the identities
\begin{equation*}
\widetilde{\l}_i(l)=\l_*+\Odr(\E^{-2ls_1(\l_*)}),\quad
\E^{-2ls_1(\widetilde{\l}_i(l))}=\Odr(\E^{-2ls_1(\l_*)}),\quad
l\to+\infty,
\end{equation*}
which are due to (\ref{1.5}), (\ref{1.6}).

Since $\boldsymbol{h}_i(l)=
\mathrm{B}^{-1}(\l_*+\tau_i^{(0)}(l),l)
\widetilde{\boldsymbol{h}}_i(l)$, Lemma~\ref{lm6.5} implies
\begin{equation*}
\boldsymbol{h}_j(l)=\sum\limits_{i=1}^{p}
\frac{\mathcal{T}_{10}^{(i)}
(\l_*+\tau_j^{(0)}(l),l)\widetilde{\boldsymbol{h}}_i(l)}{
\tau_i^{(0)}(l)+\l_*-\l_j(l)}  \boldsymbol{k}_i(l),\quad
j=1,\ldots,p.
\end{equation*}
The fractions in these identities are bounded uniformly in $\l$
and $l$ since $\boldsymbol{h}_j$ are normalized and
\begin{equation*}
Q_{ji}(l):=\frac{\mathcal{T}_{10}^{(i)}
(\l_*+\tau_j^{(0)}(l),l)\widetilde{\boldsymbol{h}}_j(l)}{
\tau_i^{(0)}(l)+\l_*-\l_i(l)} =\big(\boldsymbol{h}_j(l),
\boldsymbol{k}_i^\perp(l)\big)_{\mathbb{C}^p},
\end{equation*}
where, we remind, the vectors $\boldsymbol{k}_i^\perp$ were
introduced in the proof of Lemma~\ref{lm6.5}. We are going to
prove that the roots $\tau_j^{(0)}$ can be ordered so that the
formulas (\ref{1.11}) hold true. The matrices
\begin{equation*}
\mathrm{Q}:=
\begin{pmatrix}
Q_{11}&\ldots& Q_{1p}
\\
\vdots &  & \vdots
\\
Q_{p1}&\ldots& Q_{pp}
\end{pmatrix},
\quad \mathrm{K}=(\boldsymbol{k}_1^\perp\,\ldots\,
\boldsymbol{k}_p^\perp),\quad
\mathrm{H}=(\boldsymbol{h}_1\!\ldots\,\boldsymbol{h}_p)^t,
\end{equation*}
satisfy the identity
$\mathrm{H}\overline{\mathrm{K}}=\mathrm{Q}$. The basis
$\{\boldsymbol{k}_i^\perp\}$ being orthogonal to
$\{\boldsymbol{k}_i\}$, we conclude that
$\mathrm{K}^t=\mathrm{S}^{-1}$. Now Lemma~\ref{lm6.1} and the
condition (A) for $\mathrm{A}_0(l)$  imply the uniform in $l$
estimate $|\det Q|\geqslant C_0>0$. Hence, there exists a
permutation
\begin{equation*}
\eta=\eta(l),\quad \eta=
\begin{pmatrix}
1 &\ldots&p
\\
\eta_1&\ldots& \eta_p
\end{pmatrix}
\end{equation*}
such that
\begin{equation}\label{6.14a}
\prod\limits_{i=1}^p |Q_{i\eta_i}|\geqslant\frac{C_0}{p!} \,.
\end{equation}
This fact is proved easily by the contradiction employing the
definition of the determinant. The normalization of
$\boldsymbol{h}_i$ and the identities (\ref{5.8}) for
$\boldsymbol{k}_i^\perp$ imply that $|Q_{i\eta_i}|\leqslant
\|\boldsymbol{h}_i\|_{\mathbb{C}^p}
\|\boldsymbol{k}_{\eta_i}^\perp\|_{\mathbb{C}^p}=1$. Hence, by
(\ref{6.14a}) we obtain
\begin{equation*}
\frac{C_0}{|Q_{j\eta_j}|p!}\leqslant \prod\limits_{
\genfrac{}{}{0 pt}{}{i=1}{i\not=j}}^p |Q_{i\eta_i}|\leqslant
1,\quad \frac{C_0}{p!}\leqslant |Q_{j\eta_j}|.
\end{equation*}
Substituting the definition of $Q_{i\eta_i}$ in this estimate,
we have
\begin{align*}
|\l_i-\l_*-\tau_i^{(0)}|&\leqslant
C\left|\mathcal{T}_{10}^{(i)}(\l_*+\tau_{\eta_i}^{(0)},l)
\widetilde{\boldsymbol{h}}_i\right|\leqslant
C\big(\|\mathrm{A}_1\|+|\tau_{\eta_i}^{(0)}
|l^2\E^{-4ls_1(\l_*)}\big).
\end{align*}
Here we have also used the boundedness of
$\mathcal{T}_{10}^{(i)}$ (see Lemma~\ref{lm6.5}) and
(\ref{6.16}). Rearranging the roots $\tau_i^{(0)}$ as
$\tau_i^{(0)}:=\tau_{\eta_i}^{(0)}$, we complete the proof.
\end{proof}

\begin{proof}[Proof of Theorem~\ref{th1.6}]
Let us prove first that the matrix $\mathrm{A}(\l_*,l)$ obeys
the representation (\ref{1.10}), where all the elements of
$\mathrm{A}_0$ are zero except ones standing on the intersection
of the first row and of $(p_-+1)$-th column and $(p_-+1)$-th
column and the first row, and these elements are given by
\begin{equation*}
A_{0,1p_-+1}(l)=\overline{A}_{0,p_-+11}(l)=
2s_1(\l_*)\overline{\b}_-\b_+\E^{-2ls_1(\l_*) }.
\end{equation*}
Also we are going to prove that the corresponding matrix
$\mathrm{A}_1$ satisfies the estimate
\begin{equation}\label{1.16}
\|\mathrm{A}_1(l)\|=\Odr(l\E^{-4l\sqrt{\nu_1-\l_*}}),\quad
l\to+\infty.
\end{equation}

The definition of $\Phi_i$, the formulas (\ref{3.8a}) for
$\vp_j^\pm$, (\ref{5.1}), (\ref{4.12}) imply that
\begin{equation}\label{6.22}
\boldsymbol{\Phi}_i(\cdot,\l_*,l)=\boldsymbol{\phi}_i(\cdot,l)+
\Odr\big(l\E^{-4ls_1(\l_*)}\big), \quad l\to+\infty,
\end{equation}
in $L_2(\Om_-)\oplus L_2(\Om_+)$-norm. The identity (\ref{6.22})
and the definition of $A_{ij}$ and $\mathcal{T}_1^{(i)}$ yield
that for $i,j=1,\ldots,p_-$
\begin{equation}\label{6.24}
A_{ij}(\l_*,l)=\mathcal{T}_1^{(i)}\Phi_j(\cdot,\l_*,l)=
\mathcal{T}_1^{(i)}\phi_j(\cdot,l)+\Odr\big(l\E^{-4ls_1(\l_*)}\big
)= \Odr\big(l\E^{-4ls_1(\l_*)}\big),
\end{equation}
since $\mathcal{T}_1^{(i)}\phi_j(\cdot,l)=0$,
$i,j=1,\ldots,p_-$. In the same way one can check easily the
same identity for $i,j=p_-+1,\ldots,p$. Taking into account the
definition (\ref{3.8a}) of $\vp_1^-$, and (\ref{1.14}) by direct
calculations we obtain that
\begin{align*}
&A_{1p_-+1}(\l_*,l)=\big(\psi_1^+,
\vp_1^-(\cdot,l)\big)_{L_2(\Om_+)}=\big(\psi_1^+,\mathcal{L}_+
\mathcal{S}(2l)\psi_1^-\big)_{L_2(\Om_+)}=
\\
&=\overline{\b}_-\E^{-2ls_1(\l_*)}\big(\psi_1^+,\mathcal{L}_+
\E^{-s_1(\l_*)x_1}\phi_1\big)_{L_2(\Om_+)}+
\Odr\big(\E^{-2(s_1l(\l_*)+s_2(\l_*))}\big),\quad l\to+\infty.
\end{align*}
Now we use the definition of $\mathcal{L}_+$, (\ref{1.14}), and
integrate by parts,
\begin{equation}\label{6.24a}
\begin{aligned}
&\big(\psi_1^+,\mathcal{L}_+\E^{-s_1(\l_*)x_1}
\phi_1\big)_{L_2(\Om_+)}=\big(\psi_1^+,(-\D-\l_*+
\mathcal{L}_+)\E^{-s_1(\l_*)x_1} \phi_1\big)_{L_2(\Pi)}
\\
&=\lim\limits_{x_1\to-\infty} \left(
\left(\E^{-s_1(\l_*)x_1}\overline{\phi}_1,\frac{\p\psi_1^+}{\p
x_1}\right)_{L_2(\om)}-\left(\psi_1^+,\frac{\p}{\p
x_1}\E^{-s_1(\l_*)x_1}\overline{\phi}_1\right)_{L_2(\om)}\right)
\\
&=2s_1(\l_*)\b_+,
\end{aligned}
\end{equation}
which together with previous formula implies
\begin{equation}\label{6.25}
A_{1p_-+1}(\l_*,l)=2\overline{\b}_-\b_+s_1(\l_*)
\E^{-2ls_1(\l_*)}+
\Odr\big(\E^{-2l(s_1(\l_*)+s_2(\l_*))}\big),\quad l\to+\infty.
\end{equation}
In the same way one can show that
\begin{align*}
&A_{1j}(\l_*,l)=\Odr\big(\E^{-2l(s_1(\l_*)+s_2(\l_*))}\big),\quad
j=p_-+2,\ldots,p,
\\
&A_{ij}(\l_*,l)=\Odr\big(\E^{-4ls_2(\l_*)}\big),\quad
i=2,\ldots,p_-,\quad j=p_-+1,\ldots,p,
\\
&A_{p_-+11}(\l_*,l)=2\b_-\overline{\b}_+s_1(\l_*)
\E^{-2ls_1(\l_*)}+ \Odr\big(\E^{-2l(s_1(\l_*)+s_2(\l_*))}\big),
\\
&A_{p_-+1j}(\l_*,l)=\Odr\big(\E^{-2l(s_1(\l_*)+s_2(\l_*))}\big),
\quad j=2,\ldots,p_-,
\\
&A_{ij}(\l_*,l)=\Odr\big(\E^{-4ls_2(\l_*)}\big),\quad
i=p_-+2,\ldots,p,\quad j=1,\ldots,p_-,
\end{align*}
as $l\to+\infty$. The formulas obtained and (\ref{6.24}),
(\ref{6.25}) lead us to the representation (\ref{1.10}), where
the matrix $\mathrm{A}_0$ is defined as described, while the
matrix $\mathrm{A}_1$ satisfies (\ref{1.16}). The matrix
$\mathrm{A}_0$ being hermitian, it satisfies the condition (A).
This fact can be proved completely by analogy with
Lemma~\ref{lm6.1}.

Let us calculate the roots of $\det(\tau
\mathrm{E}-\mathrm{A}_0)$. For the sake of brevity within the
proof we denote
$c:=-2\overline{\b}_-\b_+s_1(\l_*)\E^{-2ls_1(\l_*)}$. Expanding
the determinant $\det(\tau \mathrm{E}-\mathrm{A}_0)$ w.r.t. the
first column, one can make sure that
\begin{equation*}
\det(\tau \mathrm{E}-\mathrm{A}_0)= \tau^{p-2}(\tau^2-|c|^2).
\end{equation*}
Thus, $\tau_0^{(1)}(l)=\ldots=\tau_0^{(p-2)}(l)=0$ is a root of
multiplicity $(p-2)$, and $\tau_0^{(p-1)}(l)=-2|c|$,
$\tau_0^{(p)}(l)=2|c|$. Applying Theorem~\ref{th1.5}, we arrive
at (\ref{1.17}).
\end{proof}

\begin{proof}[Proof of Theorem~\ref{th1.7}]
As in the proof of Theorem~\ref{th1.6}, we begin with the
proving (\ref{1.10}). Namely, we are going to show that
\begin{equation}\label{6.26}
\mathrm{A}_0(l)=\diag\{A_{11}(\l_*,l),0,\ldots,0\}, \quad
\|\mathrm{A}_1(l)\|=\Odr\big(\E^{-2l(s_1(\l_*)+s_2(\l_*))}\big).
\end{equation}

The definition of $\mathrm{A}$ implies
\begin{equation}\label{6.28}
A_{ij}(\l_*,l)=-\big(\mathcal{T}_6^-(\l_*,l)(\I-
\mathcal{T}_7^+(\l_*,l)\mathcal{T}_6^-(\l_*,l))^{-1}
\vp_i^-(\cdot,l),\psi_j^- \big)_{L_2(\Pi)}.
\end{equation}
Employing the definition (\ref{3.6a}) of $\mathcal{T}_6^-$, for
any $f\in L_2(\Pi,\Om_+)$ we check that
\begin{align*}
&\big(\mathcal{T}_6^-(\l_*,l)f,\psi_j^-\big)_{L_2(\Pi)}=
\big(\mathcal{L}_-\mathcal{S}(-2l)
(\mathcal{H}_+-\l_*)^{-1}f,\psi_j^-\big)_{L_2(\Pi)}
\\
&= \big(\mathcal{S}(-2l)
(\mathcal{H}_+-\l_*)^{-1}f,\mathcal{L}_-\psi_j^-\big)_{L_2(\Pi)}=
\big( (\mathcal{H}_+-\l_*)^{-1}f,\mathcal{S}(2l)(\D+\l_*)
\psi_j^-\big)_{L_2(\Pi)}
\\
&=-\big((\mathcal{H}_+-\l_*)^{-1}f,(\mathcal{H}_+-
\l_*-\mathcal{L}_+)\mathcal{S}(2l) \psi_j^-\big)_{L_2(\Pi)}
\\
&=\big(f,(\mathcal{H}_+-\l_*)^{-1}\mathcal{L}_+
\mathcal{S}(2l)\psi_j^--\mathcal{S}(2l)\psi_j^-\big)_{L_2(\Pi)}.
\end{align*}
Using this identity, (\ref{6.28}), (\ref{3.7}), (\ref{3.9}), and
Lemma~\ref{lm1.1}, we obtain that for $(i,j)\not=(1,1)$ the
functions $A_{ij}$ satisfy the relation
\begin{align*}
|A_{ij}|&\leqslant \|(\I- \mathcal{T}_7^+\mathcal{T}_6^-)^{-1}
\vp_i^-\|_{L_2(\Om_+)}\|(\mathcal{H}_+-\l_*)^{-1}\mathcal{L}_+
\mathcal{S}(2l)\psi_j^--\mathcal{S}(2l)\psi_j^-\|_{L_2(\Om_+)}=
\\
&=\Odr\big(\E^{-2l(s_1(\l_*)+s_2(\l_*))}\big),\quad l\to+\infty,
\end{align*}
where in the arguments $\l=\l_*$, and the operator
$(\mathcal{H}_+-\l_*)^{-1}$ is bounded since $p_+=0$. The
identities (\ref{6.26}) therefore hold true.

We apply now Theorem~\ref{th1.5} and infer that the formulas
(\ref{1.18}) are valid for the eigenvalues $\l_i$,
$i=1,\ldots,p-1$, while the eigenvalue $\l_p$ satisfies
\begin{equation}\label{6.30}
\l_p(l)=\l_*+A_{11}(\l_*,l)\big(1+\Odr(l^2\E^{-2ls_1(\l_*)})\big)
+ \Odr\big(\E^{-2l(s_1(\l_*)+s_2(\l_*))}\big),\quad l\to+\infty.
\end{equation}
Now it is sufficient to find out the asymptotic behaviour of
$A_{11}(\l_*,l)$.

The formula (\ref{6.28}) for $A_{11}$, Lemma~\ref{lm1.1}, and
(\ref{3.7}), (\ref{3.9}) yield
\begin{align*}
&A_{11}(\l_*,l)=-\big(\mathcal{T}_6^-(\l_*,l)\vp_1^-(\cdot,l),
\psi_1^-\big)_{L_2(\Pi)}+\Odr\big(l\E^{-8ls_1(\l_*)}\big)
\\
&=\b_-\E^{-2ls_1(\l_*)}
\big(\mathcal{T}_6^-(\l_*,l)\mathcal{L}_+\E^{-s_1(\l_*)x_1},
\psi_1^-\big)_{L_2(\Pi)}
+\Odr\big(\E^{-2l(s_1(\l_*)+s_2(\l_*))}+l\E^{-8ls_1(\l_*)}\big),
\\
&\big(\mathcal{T}_6^-(\l_*,l)\mathcal{L}_+\E^{-s_1(\l_*)x_1},
\psi_1^-\big)_{L_2(\Pi)}=\big(\mathcal{L}_-\mathcal{S}(-2l)
(\mathcal{H}_+-\l_*)^{-1}\mathcal{L}_+\E^{-s_1(\l_*)x_1},
\psi_1^-\big)_{L_2(\Om_-)}
\\
&=\big(\mathcal{S}(-2l)
(\mathcal{H}_+-\l_*)^{-1}\mathcal{L}_+\E^{-s_1(\l_*)x_1},
\mathcal{L}_-\psi_1^-\big)_{L_2(\Om_-)}= \big(\mathcal{S}(-2l)
U_+,\mathcal{L}_-\psi_1^-\big)_{L_2(\Om_-)}.
\end{align*}
Lemma~\ref{lm3.1} implies that the function $U$ satisfies
(\ref{1.19}) that determines the constant $\widetilde{\b}_-$
uniquely. Employing (\ref{1.19}), we continue our calculations,
\begin{equation*}
\big(\mathcal{S}(-2l) U_+,
\mathcal{L}_-\psi_1^-\big)_{L_2(\Om_-)}=\widetilde{\b}_-
\E^{-2ls_1(\l_*)}\big(\E^{s_1(\l_*)x_1}\phi_1(x'),
\mathcal{L}_-\psi_1^-\big)_{L_2(\Pi)}+\Odr(\E^{-2ls_2(\l_*)}).
\end{equation*}
Integrating by parts in the same way as in (\ref{6.24a}), we
obtain
\begin{equation*}
\big(\E^{s_1(\l_*)x_1}\phi_1(x'),
\mathcal{L}_-\psi_1^-\big)_{L_2(\Pi)}=\big(\mathcal{L}_-
\E^{s_1(\l_*)x_1}\phi_1(x'),
\psi_1^-\big)_{L_2(\Pi)}=-2s_1(\l_*)\overline{\b}_-.
\end{equation*}
Therefore,
\begin{equation*}
A_{11}(\l_*,l)=2s_1(\l_*)|\b_-|^2\widetilde{\b}_-\E^{-4ls_1(\l_*)}
+\Odr\big(\E^{-2l(s_1(\l_*)+
s_2(\l_*))}+l\E^{-8ls_1(\l_*)}\big).
\end{equation*}
Substituting this identity into (\ref{6.30}), we arrive at the
required formula for $\l_p$.
\end{proof}

The proof of Theorem~\ref{th1.8} is completely analogous to that
of Theorem~\ref{th1.7}.

\section{Examples}

In this section we provide some examples of the operators
$\mathcal{L}_\pm$. In what follows we will often omit the index
''$\pm$'' in the notation $\mathcal{L}_\pm$, $\mathcal{H}_\pm$,
$\Om_\pm$, $a_\pm$, writing $\mathcal{L}$, $\mathcal{H}$, $\Om$,
$a$ instead.

\textbf{1. Potential.} The simplest example is the
multiplication operator $\mathcal{L}=V$, where $V=V(x)\in
C(\overline{\Pi})$ is a real-valued compactly supported
function. Although this example is classical one for the
problems in the whole space, to our knowledge, the double-well
problem in waveguide has not been considered yet.

\textbf{2. Second order differential operator.} This is a
generalization of the previous example. We introduce the
operator $\mathcal{L}$ as
\begin{equation}\label{7.1}
\mathcal{L}=\sum\limits_{i,j=1}^{n}b_{ij}\frac{\p^2}{\p x_i\p
x_j}+\sum\limits_{i=1}^{n} b_i\frac{\p}{\p x_i}+b_0,
\end{equation}
where the complex-valued functions $b_{ij}=b_{ij}(x)$ are
piecewise continuously differentiable in $\overline{\Pi}$,
$b_i=b_i(x)$ are complex-valued functions piecewise continuous
in $\overline{\Pi}$. These functions are assumed to be compactly
supported. The only restriction to the functions are the
conditions (\ref{1.1}), (\ref{1.2}); the self-adjointness of
$\mathcal{H}$ and $\mathcal{H}_l$ is implied by these
conditions. One of the possible way to choose the functions in
(\ref{7.1}) is as follows
\begin{equation}\label{7.2}
\mathcal{L}=\dvr \mathrm{G}\nabla+\iu\sum\limits_{i=1}^{n}
\left(b_i\frac{\p}{\p x_i}-\frac{\p}{\p x_i}b_i\right)+b_0,
\end{equation}
where $\mathrm{G}=\mathrm{G}(x)$ is $n\times n$ hermitian matrix
with piecewise continuously differentiable coefficients,
$b_i=b_i(x)$ are real-valued piecewise continuously
differentiable functions, $b_0=b_0(x)$ is a real-valued
piecewise continuous function. The matrix $\mathrm{G}$ and the
functions $b_i$ are assumed to be compactly supported and
\begin{equation*}
(\mathrm{G}(x)y,y)_{\mathbb{C}^n}\geqslant
-c_0\|y\|^2_{\mathbb{C}^n}, \qquad x\in\overline{\Pi},\quad
y\in\mathbb{C}^n.
\end{equation*}
The constant $c_0$ is independent of $x$, $y$ and satisfies
(\ref{1.3}). The matrix $\mathrm{G}$ is not necessarily
non-zero. In the case $\mathrm{G}=0$ one has an example of a
first order differential operator.

\textbf{3. Magnetic Schr\"odinger operator.} This is the example
with a compactly supported magnetic field. The operator
$\mathcal{L}$ is given by  (\ref{7.2}), where $\mathrm{G}=0$.
The coefficients $b_j$ form a magnetic real-valued
vector-potential $\boldsymbol{b}=(b_1,\ldots,b_n)\in
C^1(\overline{\Pi})$, and
$b_0=\|\boldsymbol{b}\|_{\mathbb{C}^n}^2+V$, where $V=V(x)\in
C(\overline{\Pi})$ is a compactly supported real-valued electric
potential. The main assumption is the identities
\begin{equation}\label{7.3}
\frac{\p b_j}{\p x_i}=\frac{\p b_i}{\p x_j}, \qquad
x\in\overline{\Pi}\setminus\Om,\quad i,j=1,\ldots,n.
\end{equation}
To satisfy the conditions required for $\mathcal{L}$, the
magnetic vector potential should have a compact support. We are
going to show that one can always achieve it by employing the
gauge invariance.

The operator $-\D^{(D)}+\mathcal{L}$ can be represented as
$-\D^{(D)}+\mathcal{L}=(\iu\nabla+\boldsymbol{b})^2+V$, and for
any $\beta=\beta(x)\in C^2(\overline{\Pi})$ the identity
\begin{equation*}
\E^{-\iu\b}(\iu\nabla+\boldsymbol{b})^2\E^{\iu\b}=(\iu\nabla+
\boldsymbol{b}-\nabla\b)^2
\end{equation*}
holds true. In view of (\ref{7.3}) we conclude that there exist
two functions $\b_\pm=\b_\pm(x)$ belonging to
$C^2(\overline{\Pi\cap\{x: \pm x_1>a\}})$ such that
$\nabla\b_\pm=\boldsymbol{b},\quad x\in\Pi\cap\{x: \pm x_1>a\}$.
We introduce now the function $\b$ as
\begin{equation*}
\b(x)=
\begin{cases}
\chi(a+x_1+1)\b_-(x), & x_1\in(-\infty,-a),\quad x'\in\om,
\\
0, & x_1\in[-a,a],\quad x'\in\om,
\\
\chi(a-x_1+1)\b_+(x), & x_1\in(a,+\infty),\quad x'\in\om,
\end{cases}
\end{equation*}
where, we remind, the cut-off function $\chi$ was introduced in
the proof of Lemma~\ref{lm2.1}. Clearly, $\b\in
C^2(\overline{\Pi})$, and $\nabla\b=\boldsymbol{b}$,
$x\in\overline{\Pi}\setminus\{x: |x_1|<a+1\}$. Therefore, the
vector $\boldsymbol{b}-\nabla\b$ has compact support.

If one of the distant perturbations in the operator
$\mathcal{H}_l$, say, the right one, is a compactly supported
magnetic field, it is sufficient to employ the gauge
transformation $\psi(x)\mapsto
\E^{\iu\b(x_1-l,x')}\widetilde{\psi}(x)$ to satisfy the
conditions for $\mathcal{L}_+$.

\begin{figure}[t]
\begin{center}
\noindent
\includegraphics
[width=8.26 true cm, height=2.72 true cm, ]{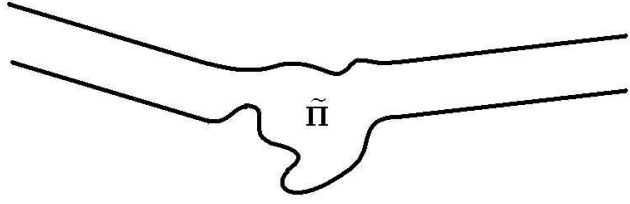}
\end{center}
\caption{Geometric perturbation}\label{fig1}
\end{figure}

\textbf{4. Curved and deformed waveguide.} One more interesting
example is a geometric perturbation. Quite popular cases are
local deformation of the boundary and curving the waveguide
(see, for instance, \cite{CDFK}, \cite{BGRS}, \cite{BEGK}, and
references therein). Here we consider the case of general
geometric perturbation, which includes in particular deformation
and curving. Namely, let $\widetilde{x}=\mathcal{G}(x)\in
C^2(\overline{\Pi})$ be a diffeomorphism, where
$\widetilde{x}=(\widetilde{x}_1,\ldots,\widetilde{x}_n)$,
$\mathcal{G}(x)=(\mathcal{G}_1(x),\ldots,\mathcal{G}_n(x))$. We
denote $\widetilde{\Pi}:=\mathcal{G}(\Pi)$,
$\mathcal{P}:=\mathcal{G}^{-1}$. By $\mathrm{P}=\mathrm{P}(x)$
we indicate the matrix
\begin{equation*}
\mathrm{P}:=
\begin{pmatrix}
\frac{\p \mathcal{P}_1}{\p \widetilde{x}_1} & \ldots & \frac{\p
\mathcal{P}_1}{\p \widetilde{x}_n}
\\
\vdots & & \vdots
\\
\frac{\p \mathcal{P}_n}{\p \widetilde{x}_1} & \ldots & \frac{\p
\mathcal{P}_n}{\p \widetilde{x}_n}
\end{pmatrix},
\end{equation*}
while the symbol $\mathrm{p}=\mathrm{p}(x)$ denotes the
corresponding Jacobian, $\mathrm{p}(x):=\det \mathrm{P}(x)$. The
function $\mathrm{p}(x)$ is supposed to have no zeroes in
$\overline{\Pi}$. The main assumption we make is
\begin{equation}\label{7.4}
\mathrm{P}(x)=const, \qquad \mathrm{P}^t\mathrm{P}=\mathrm{E},
\quad x\in\overline{\Pi}\setminus\Om.
\end{equation}
It implies that outside $\Om$ the mapping $\mathcal{P}$ acts as
a combination of a shift and a rotating. Hence, the part of
$\widetilde{\Pi}$ given by $\mathcal{G}((a,+\infty)\times\om)$
is also a tubular domain being a direct product of a half-line
and $\om$. The same is true for
$\mathcal{G}((-\infty,-a)\times\om)$. The typical example of the
domain $\widetilde{\Pi}$ is given on Figure~\ref{fig1}.

Let $-\widetilde{\D}^{(D)}$ be the negative Dirichlet Laplacian
in $L_2(\widetilde{\Pi})$ with the domain
$\Ho^2(\widetilde{\Pi})$. It is easy to check that the operator
$\mathcal{U}:L_2(\widetilde{\Pi})\to L_2(\Pi)$ defined as
\begin{equation}\label{7.5}
(\mathcal{U}v)(x):=\mathrm{p}^{-1/2}(x)v(\mathcal{P}^{-1}(x))
\end{equation}
is unitary. The operator
$\mathcal{H}:=-\mathcal{U}\widetilde{\D}^{(D)}\mathcal{U}^{-1}$
has $\Ho^2(\Pi)$ as the domain and is self-adjoint in
$L_2(\Pi)$. It can be represented as
$\mathcal{H}=-\D^{(D)}+\mathcal{L}$, where $\mathcal{L}$ is a
second order differential operator
\begin{equation}\label{7.6}
\mathcal{L}=-\mathrm{p}^{1/2}\dvr_x
\mathrm{p}^{-1}\mathrm{P}^t\mathrm{P}\nabla_x
\mathrm{p}^{1/2}+\D_x.
\end{equation}
Indeed, for any $u_1,u_2\in C_0^\infty(\Pi)$
\begin{equation}
\begin{aligned}
&(\mathcal{H}u_1,u_2)_{L_2(\Pi)}=-(\widetilde{\D}^{(D)}
\mathcal{U}^{-1}u_1,\mathcal{U}^{-1}u_2)_{L_2(\widetilde{\Pi})}
=\int\limits_{\widetilde{\Pi}} \left(\nabla_{\widetilde{x}}
\mathrm{p}^{1/2}u_1,\nabla_{\widetilde{x}}
\mathrm{p}^{1/2}u_2\right)_{\mathbb{C}^n} \di \widetilde{x}=
\\
&=\int\limits_{\Pi} \left(\mathrm{P}\nabla_{x}
\mathrm{p}^{1/2}u_1,\mathrm{p}^{-1}\mathrm{P}\nabla_{x}
\mathrm{p}^{1/2}u_2\right)_{\mathbb{C}^n} \di x= -
\int\limits_{\Pi} \overline{u}_2\mathrm{p}^{1/2}\dvr_x
\mathrm{p}^{-1}\mathrm{P}^t\mathrm{P}\nabla_{x}
\mathrm{p}^{1/2}u_1\di x.
\end{aligned}\label{7.7}
\end{equation}
The assumption (\ref{7.4}) yields that $\mathrm{p}=1$ holds for
$x\in\overline{\Pi}\setminus\Om$, and therefore the coefficients
of the operator $\mathcal{L}$ have the support inside
$\overline{\Om}$.

We are going to check the conditions (\ref{1.1}), (\ref{1.2}),
(\ref{1.3}) for the operator $\mathcal{L}$ introduced by
(\ref{7.6}). The symmetricity is obvious, while the estimates
follow from (\ref{7.6}), (\ref{7.7}),
\begin{align*}
(\mathcal{L}u&,u)_{L_2(\Om)}=
\|\mathrm{p}^{-1/2}\mathrm{P}\nabla_x \mathrm{p}^{1/2}
u\|_{L_2(\Om)}^2-\|\nabla u\|_{L_2(\Om)}^2\geqslant C\|\nabla_x
\mathrm{p}^{1/2} u\|_{L_2(\Om)}^2-\|\nabla u\|_{L_2(\Om)}^2
\\
&=C\left(\|\mathrm{p}^{1/2} \nabla_x
u\|_{L_2(\Om)}^2+\|u\nabla\mathrm{p}^{1/2}\|_{L_2(\Om)}^2 +
2(\mathrm{p}^{1/2} \nabla_x
u,u\nabla\mathrm{p}^{1/2})_{L_2(\Om)}\right) -\|\nabla
u\|_{L_2(\Om)}^2
\\
&\geqslant C\left(\frac{1}{2}\|\mathrm{p}^{1/2} \nabla_x
u\|_{L_2(\Om)}^2-\|u\nabla\mathrm{p}^{1/2}\|_{L_2(\Om)}^2\right)
-\|\nabla u\|_{L_2(\Om)}^2
\\
&\geqslant -\left(1-\frac{C}{2}\right)\| \nabla_x
u\|_{L_2(\Om)}^2-C\|u\|_{L_2(\Om)}^2,
\end{align*}
where $C>0$ is a constant.

If both the operators $\mathcal{L}_\pm$ are the geometric
perturbations described by the diffeomorphisms
$\mathcal{G}^\pm(x)$, without loss of generality we can assume
that $\mathcal{G}^\pm(x)\equiv x$ as $\pm x_1<-a_\pm$. We
introduce now one more diffeomorphism
\begin{equation*}
\mathcal{G}_l(x):=
\begin{cases}
\mathcal{G}^+(x_1-l,x')+(l,0,\ldots,0), & x_1\geqslant 0,
\\
\mathcal{G}^-(x_1+l,x')-(l,0,\ldots,0), & x_1\leqslant 0,
\end{cases}
\end{equation*}
This mapping is well-defined for $l>\max\{a_-,a_+\}$. The domain
$\widetilde{\Pi}_l:=\mathcal{G}_l(\Pi)$ can be naturally
regarded as a waveguide with two distant geometric
perturbations. Considering the negative Dirichlet Laplacian in
$L_2(\widetilde{\Pi})$, we obtain easily an unitary equivalent
operator in $L_2(\Pi)$. The corresponding unitary operator is
defined by (\ref{7.5}) with $\mathcal{P}$ replaced by
$\mathcal{G}_l^{-1}$ and the similar replacement for
$\mathrm{p}$ is required. Clearly, the obtained operator in
$L_2(\Pi)$ is the operator $\mathcal{H}_l$ generated by the
operators $\mathcal{L}_\pm$ associated with $\mathcal{G}^\pm$.

\textbf{5. Delta interaction.} Our next example is the delta
interaction supported by a manifold. Let $\G$ be a closed
bounded $C^3$-manifold in $\Pi$ of the codimension one and
oriented by a normal vector field $\boldsymbol{\nu}$. It is
supposed that $\G\cap\p\Pi=\emptyset$. The manifold $\G$ can
consist of several components. By $\xi=(\xi_1,\ldots,\xi_{n-1})$
we denote coordinates on $\G$, while $\widetilde{\vr}$ will
indicate the distance from a point to $\G$ measured in the
direction of $\boldsymbol{\nu}=\boldsymbol{\nu}(\xi)$. We assume
that $\G$ is so that the coordinates $(\widetilde{\vr},\xi)$ are
well-defined in a neighbourhood of $\G$. Namely, we suppose that
the mapping $(\vr,\xi)=\mathcal{P}_\G(\widetilde{x})$ is
$C^3$-diffeomorphism, where
$\widetilde{x}=(\widetilde{x}_1,\ldots,\widetilde{x}_n)$ are the
coordinates in $\Pi$. Let $b=b(\xi)\in C^3(\G)$ be a real-valued
function. The operator in question is the negative Laplacian
defined on the functions
$v\in\H^2(\Pi\setminus\G,\p\Pi)\cap\Ho^1(\Pi)$ satisfying the
condition
\begin{equation}\label{7.9}
v\big|_{\widetilde{\vr}=+0}-v\big|_{\widetilde{\vr}=-0}=0,\quad
\frac{\p v}{\p\widetilde{\vr}}\Big|_{\widetilde{\vr}=+0}-
\frac{\p v}{\p\widetilde{\vr}}\Big|_{\widetilde{\vr}=-0}=
bv\big|_{\widetilde{\vr}=0}.
\end{equation}
We indicate this operator as $\mathcal{H}_\G$. An alternative
way to introduce $\mathcal{H}_\G$ is via associated quadratic
form
\begin{equation}\label{7.10b}
\mathfrak{q}_\G(v_1,v_2):=(\nabla_{\widetilde{x}}v_1,
\nabla_{\widetilde{x}}v_2)_{L_2(\Pi)}+(bv_1,v_2)_{L_2(\G)},
\end{equation}
where $u\in\Ho^1(\Pi)$ (see, for instance, \cite[Appendix K,
Sec. K.4.1]{AGHH}, \cite[Remark 4.1]{BEKS}). It is known that
the operator $\mathcal{H}_\G$ is self-adjoint.

We are going to show that there exists a diffeomorpism
$x=\mathcal{P}(\widetilde{x})$ such that a unitary operator
$\mathcal{U}$ defined by (\ref{7.5}) maps $L_2(\Pi)$ onto
$L_2(\Pi)$, and the operator
$\mathcal{U}\mathcal{H}_\G\mathcal{U}^{-1}$ has $\Ho^2(\Pi)$ as
the domain. We will also show that
$\mathcal{U}\mathcal{H}_\G\mathcal{U}^{-1}=-\D^{(D)}+\mathcal{L}$,
where the operator $\mathcal{L}$ is given by (\ref{7.1}).

First we introduce an auxiliary mapping as
\begin{equation*}
\widetilde{\mathcal{P}}(\widetilde{x}):=
\mathcal{P}_\G^{-1}\left(\vr,\xi\right),\quad
\vr:=\widetilde{\vr}+
\frac{1}{2}\widetilde{\vr}|\widetilde{\vr}|b(\xi),\quad
(\widetilde{\vr},\xi)=\mathcal{P}_\G(x).
\end{equation*}
The coordinates $(\widetilde{\vr},\xi)$ are well-defined in a
neighbourhood of $\G$, which can be described as $\{x:
|\widetilde{\vr}|<\d\}$, where $\d$ is small enough. Indeed,
\begin{equation}\label{7.10}
\widetilde{\mathcal{P}}_i(\widetilde{x})=\widetilde{x}_i+
\widetilde{\vr}|\widetilde{\vr}|
\widehat{\mathcal{P}}_i(\widetilde{x}),
\end{equation}
where the functions $\widehat{\mathcal{P}}_i(x)$ are
continuously differentiable in a neighbourhood of $\G$.
Therefore, the Jacobian of $\widetilde{\mathcal{P}}_i$ tends to
one as $\widetilde{\vr}\to0$ uniformly in $\xi\in\G$. Now we
construct the required mapping as follows
\begin{equation}\label{7.10a}
x=\mathcal{P}(\widetilde{x}):=
\big(1-\widetilde{\chi}(\widetilde{\vr}/{\d}
)\big)\widetilde{x}+ \widetilde{\chi}(\widetilde{\vr}/{\d})
\widetilde{\mathcal{P}}(\widetilde{x}).
\end{equation}
The symbol $\widetilde{\chi}(t)$ indicates an even infinitely
differentiable cut-off function being one as $|t|<1$ and
vanishing as $|t|>2$. We assume also that $\d$ is chosen so that
$\supp\widetilde{\chi}(\widetilde{\vr}/\d)\cap\p\Pi=\emptyset$.
Let us prove that $\mathcal{P}$ is a $C^1$-diffeomorhism and
$\mathcal{P}(\overline{\Pi})=\overline{\Pi}$.

It is obvious that $\mathcal{P}\in C^1(\overline{\Pi})$. If
$\widetilde{\chi}(\widetilde{\vr}/\d)=0$, the mapping
$\mathcal{P}$ acts as an identity mapping, and therefore for
such $x$ the Jacobian $\mathrm{p}=\mathrm{p}(\widetilde{x})$ of
$\mathcal{P}$ equals one. The function $\widetilde{\chi}$ is
non-zero only in a small neighbourhood $\{\widetilde{x}:
|\widetilde{\vr}|<2\d\}$ of $\G$, where the identities
(\ref{7.10}) are applicable. These identities together with the
definition of $\mathcal{P}$ imply that
\begin{equation*}
\mathrm{p}(\widetilde{x})=1+\widetilde{\vr}
\mathrm{p}_1(\widetilde{x}),
\end{equation*}
where the function $\mathrm{p}_1(\widetilde{x})$ is bounded
uniformly in $\d$ and $\widetilde{x}$ as
$|\widetilde{\vr}|\leqslant 2\d$. Hence, we can choose $\d$
small enough so that  $\mathrm{p}\geqslant 1/2$ as
$|\widetilde{\vr}|\leqslant 2\d$. Therefore, $\mathcal{P}$ is
$C^1$-diffeomorphism. As $x$ close to  $\p\Pi$, the
diffeomorphism $\mathcal{P}$ acts as the identity mapping. It
yields that $\mathcal{P}(\p\Pi)=\p\Pi$,
$\mathcal{P}(\overline{\Pi})=\overline{\Pi}$. Since $\vr=0$ as
$\widetilde{\vr}=0$, it follows that $\mathcal{P}(\G)=\G$.

The function $\widetilde{\vr}\mapsto\vr$ is continuously
differentiable and its second derivative is piecewise
continuous. Thus, the second derivatives of
$\mathcal{P}_i(\widetilde{x})$ are piecewise continuous as well.
The same is true for the inverse mapping $\mathcal{P}^{-1}$.

We introduce the unitary operator $\mathcal{U}$ by (\ref{7.5}),
where $\mathcal{P}$ is the diffeomorphism defined by
(\ref{7.10a}). It is obvious that $\mathcal{U}$ maps $L_2(\Pi)$
onto itself. Let us prove that it maps the domain of
$\mathcal{H}_\G$ onto $\Ho^2(\Pi)$. In order to do it, we have
to study the behaviour of $\mathrm{p}$ in a vicinity of $\G$. It
is clear that $\mathrm{p}(\widetilde{x})\in
C^2(\overline{\Pi}\setminus\G)\cap C(\overline{\Pi})$, and this
function has discontinuities at $\G$ only. By
$\mathrm{p}_\G=\mathrm{p}_\G(\widetilde{\vr},\xi)$ we denote the
Jacobian corresponding to $\mathcal{P}_\G$. It is obvious that
$\mathrm{p}_\G\in C^2(\{(\widetilde{\vr},\xi):
|\widetilde{\vr}|\leqslant\d\})$. Employing the well-known
properties of the Jacobians, we can express $\mathrm{p}$ in
terms of $\mathrm{p}_\G$,
\begin{equation*}
\mathrm{p}(\widetilde{x})= \frac{(1+|\widetilde{\vr}|b(\xi))
\mathrm{p}_\G(\widetilde{\vr},\xi)}{\mathrm{p}_\G(\vr,\xi)}
=\frac{\mathrm{p}_\G(\widetilde{\vr},\xi)
(1+|\widetilde{\vr}|b(\xi))} {\mathrm{p}_\G(\widetilde{\vr}+
\frac{1}{2}\widetilde{\vr}|\widetilde{\vr}|b(\xi),\xi)} ,\quad
|\widetilde{\vr}|\leqslant\d.
\end{equation*}
This relation allows us to conclude that $\mathrm{p}\in
C^2(\{(\widetilde{\vr},\xi):
0\leqslant\widetilde{\vr}\leqslant\d\})$, $\mathrm{p}\in
C^2(\{(\widetilde{\vr},\xi):
-\d\geqslant\widetilde{\vr}\geqslant 0\})$, and
\begin{equation}\label{7.12}
\mathrm{p}^{1/2}\big|_{\widetilde{\vr}=+0}-
\mathrm{p}^{1/2}\big|_{\widetilde{\vr}=-0}=0,\quad
\frac{\p}{\p\widetilde{\vr}}\mathrm{p}^{1/2}\big|_{\vr=+0}-
\frac{\p}{\p\widetilde{\vr}} \mathrm{p}^{1/2}\big|_{\vr=-0}=b.
\end{equation}
Given $u=u(x)\in\Ho^2(\Pi)$, we introduce the function
$v=v(\widetilde{x}):=(\mathcal{U}^{-1}u)(\widetilde{x})=
\mathrm{p}^{1/2}(\widetilde{x})u(\mathcal{P}(\widetilde{x}))$.
Due to the smoothness of $\mathrm{p}$,
$v(\widetilde{x})\in\H^2(\Pi\setminus\G)\cap\Ho^1(\Pi)$. The
identities (\ref{7.12}) and the belonging $u(x)\in\H^2(\Pi)$
imply the condition (\ref{7.9}) for $v$.

Suppose now that a function
$v=v(\widetilde{x})\in\H^2(\Pi\setminus\G)\cap\Ho^1(\Pi)$
satisfies (\ref{7.9}). Due to the smoothness of $\mathcal{P}$ it
is sufficient to check that the function $\mathcal{U}v$ regarded
as depending on $\widetilde{x}$ belongs to $\Ho^2(\Pi)$, i.e.,
$u(\widetilde{x}):=\mathrm{p}^{-1/2}(\widetilde{x})
v(\widetilde{x})\in\Ho^2(\Pi)$. It is clear that
$u\in\H^2(\Pi\setminus\G)\cap\Ho^1(\Pi)$. Hence, it remains to
check the belonging $u\in\H^2(\{\widetilde{x}:
|\widetilde{\vr}|\leqslant\d\})$. The functions $\mathcal{P}_i$
being twice piecewise continuously differentiable, it is
sufficient to make sure that the function $u$ treated as
depending on $\widetilde{\vr}$ and $\xi$ is an element of
$\H^2(\{(\widetilde{\vr},\xi): |\widetilde{\vr}|<\d,
\xi\in\G\})$. We have $u\in\H^1(R)$,  $u\in\H^2(R_+)$,
$u\in\H^2(R_-)$, $R:=\{(\widetilde{\vr},\xi):
|\widetilde{\vr}|<\d, \xi\in\G\}$,
$R_\pm:=\{(\widetilde{\vr},\xi): 0<\pm\widetilde{\vr}<\pm\d,
\xi\in\G\}$. Hence, we have to prove the existence of the
generalized second derivatives for $u$ belonging to $L_2(R)$.
The condition (\ref{7.9}) and the formulas (\ref{7.12}) yield
that
\begin{equation*}
u\big|_{\widetilde{\vr}=+0}=u\big|_{\widetilde{\vr}=-0},\quad
\frac{\p u}{\p\widetilde{\vr}}\bigg|_{\widetilde{\vr}=+0}=
\frac{\p u}{\p\widetilde{\vr}}\bigg|_{\widetilde{\vr}=-0}.
\end{equation*}
Since $u\big|_{\widetilde{\vr}=\pm 0}\in\H^1(\G)$, the first of
these relations implies that
\begin{equation*}
\frac{\p u}{\p\xi_i}\bigg|_{\widetilde{\vr}=+0}= \frac{\p u
}{\p\xi_i}\bigg|_{\widetilde{\vr}=-0},\quad i=1,\ldots, n-1.
\end{equation*}
Having the obtained relations in mind, for any $\z\in C_0^2(R)$
we integrate by parts,
\begin{align*}
\left(u,\frac{\p^2\z}{\p\vr^2}\right)_{L_2(R)}&=
\left(u,\frac{\p^2\z}{\p\vr^2}\right)_{L_2(R_-)}+
\left(u,\frac{\p^2\z}{\p\vr^2}\right)_{L_2(R_+)}=
\\
&=
\left(\frac{\p^2 u}{\p\vr^2},\z\right)_{L_2(R_-)}+
\left(\frac{\p^2 u}{\p\vr^2},\z\right)_{L_2(R_+)},
\end{align*}
and in the same way we obtain:
\begin{align*}
&\left(u,\frac{\p^2\z}{\p\vr\p\xi_i}\right)_{L_2(R)}=
\left(\frac{\p^2 u}{\p\vr\p\xi_i},\z\right)_{L_2(R_-)}+
\left(\frac{\p^2 u}{\p\vr\p\xi_i},\z\right)_{L_2(R_+)},
\\
&\left(u,\frac{\p^2\z}{\p\xi_i\p\xi_j}\right)_{L_2(R)}=
\left(\frac{\p^2 u}{\p\xi_i\p\xi_j},\z\right)_{L_2(R_-)}+
\left(\frac{\p^2 u}{\p\xi_i\p\xi_j},\z\right)_{L_2(R_+)}.
\end{align*}
Thus, the generalized second derivatives of the function $u$
exist and coincide with the corresponding derivatives of $u$
regarded as an element of $\H^2(R_-)$ and $\H^2(R_+)$.

Let us show that the operator
$\mathcal{U}\mathcal{H}_\G\mathcal{U}^{-1}$ can be represented
as
$\mathcal{U}\mathcal{H}_\G\mathcal{U}^{-1}=-\D^{(D)}+\mathcal{L}$,
where $\mathcal{L}$ is given by (\ref{7.1}). Proceeding in the
same way as in (\ref{7.7}) and using (\ref{7.10b}), for any
$u_1,u_2\in C_0^\infty(\Pi)$ we obtain
\begin{equation*}
(\mathcal{U}\mathcal{H}_\G\mathcal{U}^{-1}u_1,u_2)_{L_2(\Pi)}=
\int\limits_{\Pi} \left(\nabla_{x}
\mathrm{p}^{1/2}u_1,\mathrm{p}^{-1}\mathrm{P}^t\mathrm{P}\nabla_{x}
\mathrm{p}^{1/2}u_2\right)_{\mathbb{C}^n} \di x+\int\limits_\G b
u_1\overline{u}_2\di\xi.
\end{equation*}
We have used here that $\mathrm{p}\equiv 1$ as $x\in\G$ and
$\mathcal{P}(\G)=\G$. Employing (\ref{7.12}) and having in mind
that $\mathcal{P}\big|_\G=\mathrm{E}$ due to (\ref{7.10}), we
integrate by parts,
\begin{align*}
&(\mathcal{U}\mathcal{H}_\G\mathcal{U}^{-1}u_1,u_2)_{L_2(\Pi)}=
-\int\limits_\G
\overline{u}_2\left(-\frac{\p}{\p\widetilde{\rho}}
(\sqrt{\mathrm{p}}u_1)\bigg|_{\widetilde{\rho}=+0}+
\frac{\p}{\p\widetilde{\rho}}
(\sqrt{\mathrm{p}}u_1)\bigg|_{\widetilde{\rho}=-0}+bu_1\right)\di\xi-
\\
&-\int\limits_{\Pi\setminus\G}\overline{u}_2
\mathrm{p}^{1/2}\dvr_x
\mathrm{p}^{-1}\mathrm{P}^t\mathrm{P}\nabla_{x}
\mathrm{p}^{1/2}u_1\di
x=-\int\limits_{\Pi\setminus\G}\overline{u}_2
\mathrm{p}^{1/2}\dvr_x
\mathrm{p}^{-1}\mathrm{P}^t\mathrm{P}\nabla_{x}
\mathrm{p}^{1/2}u_1\di x.
\end{align*}
Here $\Pi\setminus\G$ means that in a neighbourhood we partition
the domain of integration into two pieces one being located in
the set $\{x: |\widetilde{\vr}|<0\}$, while the other
corresponds to $|\widetilde{\vr}|>0$. Such partition is needed
since the first derivatives of $\mathrm{p}$ have jump at $\G$
and therefore the second derivatives of $\mathrm{p}$ are not
defined at $\G$. At the same time, the function $\mathrm{p}$ has
continuous second derivatives as $\widetilde{\vr}<0$ and these
derivatives have finite limit as $\widetilde{\vr}\to-0$. The
same is true for $\widetilde{\vr}>0$.

The matrix $\mathrm{P}$ is piecewise continuously
differentiable, and thus
$\mathcal{U}\mathcal{H}_\G\mathcal{U}^{-1}=-\mathrm{p}^{1/2}\dvr_x
\mathrm{p}^{-1}\mathrm{P}^t\mathrm{P}\nabla_x \mathrm{p}^{1/2}$,
where the second derivatives of $\mathrm{p}$ are treated in the
aforementioned sense. This operator is obviously self-adjoint.
Outside the set $\{x: |\widetilde{\vr}|<2\d\}$ the
diffeomorphism $\mathcal{P}$ acts as the identity mapping. It
yields that at such points $\mathrm{P}=\mathrm{E}$,
$\mathrm{p}=1$. Therefore, $\mathcal{L}$ is a differential
operator having compactly supported coefficients, and is a
particular case of (\ref{7.1}). It is also clear that the
operator $\mathcal{L}$ satisfies (\ref{1.1}). The inequality
(\ref{1.2}) can be checked in the same way how it was proved in
the previous subsection.

\textbf{6. Integral operator.} The operator $\mathcal{L}$ is not
necessary to be either a differential operator or reducible to a
differential one. An example is an integral operator
\begin{equation*}
\mathcal{L}=\int\limits_{\Om} L(x,y)u(y)\di y.
\end{equation*}
The kernel $L\in L_2(\Pi\times\Pi)$ is assumed to be symmetric,
i.e., $L(x,y)=\overline{L(y,x)}$. It is clear that the operator
$\mathcal{L}$ satisfies (\ref{1.1}), (\ref{1.2}). It is also
$\D^{(D)}$-compact and therefore the operators $\mathcal{H}$ and
$\mathcal{H}_l$ are self-adjoint.

In conclusion we should stress that all possible examples of
$\mathcal{L}$ are not exhausted by ones given above. For
instance, combinations of these examples are possible like
compactly supported magnetic field with delta interaction, delta
interaction in a deformed waveguide, integro-differential
operator, etc. Moreover, the operators $\mathcal{L}_-$ and
$\mathcal{L}_+$ are not necessary to be of the same nature. For
example, $\mathcal{L}_-$ can be a potential, while
$\mathcal{L}_+$ describes compactly supported magnetic field
with delta interaction.

\section*{Acknowledgments}

I am grateful to Pavel Exner who attracted my attention to the
problem studied in this article. I also thank him for
stimulating discussion and the attention he paid to this work.

The research was supported by \emph{Marie Curie International
Fellowship} within 6th European Community Framework Programm
(MIF1-CT-2005-006254). The author is also supported by the
Russian Foundation for Basic Researches (No. 07-01-00037) and by
the Czech Academy of Sciences and Ministry of Education, Youth
and Sports (LC06002). The author gratefully acknowledges the
support from Deligne 2004 Balzan prize in mathematics.

\renewcommand{\refname}

\end{document}